\documentclass[preprint,floats,12pt,superscriptaddress,nofootinbib,floatfix,showkeys]{revtex4-2}
\usepackage{natbib}
\usepackage{times} 
\usepackage{amssymb,amsmath}
\usepackage{csquotes}
\usepackage{physics}
\usepackage{bm}

\usepackage[usenames,dvipsnames]{color}
\usepackage{epstopdf}
\usepackage[pdftex]{graphicx}
\usepackage[pdftex]{hyperref}
\hypersetup{a4paper,
    pdftitle={Manuscript on reactive thin-film hydrodynamics} 
    pdfauthor={Florian Voss and Uwe Thiele},  
pdfproducer={lateX},
pdfview=FitV,       
pdfstartview=FitB,
linkcolor=blue,     
citecolor=blue,     
urlcolor=red,      
breaklinks=true,    
colorlinks=true,
citebordercolor=0 0 0,  
filebordercolor=0 0 0,
linkbordercolor=0 0 0,
menubordercolor=0 0 0,
urlbordercolor=0 0 0,
pdfhighlight=/I,
pdfborder=0 0 0,   
bookmarksopen=true,
bookmarksnumbered=true
}
\DeclareGraphicsExtensions{.jpg, .pdf, .tif, .png}
\usepackage[usenames,dvipsnames]{color}
\usepackage[normalem]{ulem}

\newcommand{\reviewchange}[1]{\textcolor{black}{#1}}
\newcommand{\tens}[1]{\mathbf{\underline{#1}}}

\usepackage{booktabs}
\usepackage{multirow}
\renewcommand{\thesection}{\arabic{section}}
\renewcommand{\thesubsection}{\thesection.\arabic{subsection}}
\renewcommand{\thesubsubsection}{\thesubsection.\arabic{subsubsection}}

\makeatletter
\renewcommand{\p@subsection}{}
\renewcommand{\p@subsubsection}{}
\renewcommand{\@seccntformat}[1]{\csname the#1\endcsname\quad}
\makeatother

\begin{document}
\count\footins = 1000 

\title{Gradient dynamics approach to reactive thin-film hydrodynamics}

\author{Florian Voss}
\email{f\_voss09@uni-muenster.de}
\thanks{ORCID ID: 0009-0003-9679-035X}
\affiliation{Institute of Theoretical Physics, University of M\"unster, Wilhelm-Klemm-Str.\ 9, 48149 M\"unster, Germany}

\author{Uwe Thiele}
\email{u.thiele@uni-muenster.de}
\homepage{http://www.uwethiele.de}
\thanks{ORCID ID: 0000-0001-7989-9271}
\affiliation{Institute of Theoretical Physics, University of M\"unster, Wilhelm-Klemm-Str.\ 9, 48149 M\"unster, Germany}
\affiliation{Center for Nonlinear Science (CeNoS), University of M\"unster, Corrensstr.\ 2, 48149 M\"unster, Germany}
\affiliation{Center for Multiscale Theory and Computation (CMTC), University of M\"unster, Corrensstr.\ 40, 48149 M\"unster, Germany}

\begin{abstract}
Wetting and dewetting dynamics of simple and complex liquids is described by kinetic equations in gradient dynamics form that incorporates the various coupled dissipative processes in a fully thermodynamically consistent manner. After briefly reviewing this, we also review how chemical reactions can be captured by a related gradient dynamics description, assuming detailed balanced mass action type kinetics. Then, we bring both aspects together and discuss mesoscopic reactive thin-film hydrodynamics illustrated by two examples, namely, models for reactive wetting and reactive surfactants. These models can describe the approach to equilibrium but may also be employed to study out-of-equilibrium \reviewchange{chemo-mechanical} dynamics. In the latter case, one breaks the gradient dynamics form by chemostatting to obtain active systems. In this way, for reactive wetting we recover running drops that are driven by chemically sustained wettability gradients and for drops covered by autocatalytic reactive surfactants we find complex forms of self-propulsion and self-excited oscillations.

\noindent
The published version of this preprint can be found under:
 
Voss, F., Thiele, U., J. Eng. Math. 149, 2 (2024). doi: 10.1007/s10665-024-10402-x
\end{abstract}

\maketitle
\newpage
\section{Introduction}
Phenomena involving reactive fluids are found across many scales and range from the formation of stars to combustion, biofilm growth and many intracellular processes \cite{GOK2020ssr, DeW2020arfm, Liberman2010, Donl2002eid, TJLT2017prl, NaVS2019fmb, BWWF2022nrp}. Often, occurring chemical reactions do not only amend existing flows, but give rise to novel phenomena that result from the interplay of reactions and hydrodynamic transport. In his seminal paper \cite{Turi1952ptrslsbs}, Turing showed that pattern formation is possible even if only simple diffusive transport exists in a reactive system. One may then naturally turn to more intricate hydrodynamic settings, for instance, the wetting of solid or liquid substrates by (other) liquids \cite{Genn1985rmp,BEIM2009rmp,CrMa2009rmp} and discuss how \reviewchange{chemical reactions influence} the occurring phenomena \cite{Pism1984jcis,DaPi1984jcis,BaWh1989l,MeMi1993pa,BaBM1994n,Genn1997el,ZhWT1998pre,SaCT2000am,NiAS2006pf,Hanc2011ptrsbs}, also see reviews \cite{KuPr2007aci,ShBM2010e,EuVo2016jms}.

The statics and dynamics of wetting inherently involves the study of interacting interfaces and three-phase contact lines which gives rise to a myriad of fascinating phenomena like drop spreading and dewetting of thin liquid films \cite{BEIM2009rmp,CrMa2009rmp}, contact line motion and instabilities \cite{CrMa2009rmp,SnAn2013arfm}, the lotus effect \cite{KoBB2008sm} as well as tears of wine \cite{Neog1985jcis} and surfactant-induced spreading instabilities \cite{CrMa2009rmp}. One may then expect that the additional presence of chemical reactions will greatly enrich the spectrum of interfacial effects. Indeed, examples include chemically driven running droplets \cite{BaBM1994n,DoOn1995prl,BrGe1995crassi,Genn1998pa,LeLa2000jacs,LeKL2002pre,ThJB2004prl,SKYN2005pre,JoBT2005epje,SuMY2006ptps,Kita2006ptps,YBJZ2012sm,Arsc2016l}, and chemically driven Marangoni flows that cause oscillatory bulk convective motion \cite{DMMB1996prl, MKHB2016arcmp, Mich2023arfm}, surface waves \cite{Pism1997prl,PTTK2007pf}, fingering instabilities \cite{RCBP2012cpl}, and crawling vesicular aggregates \cite{NFK2016msde}.
Such reactive \reviewchange{interface} phenomena are highly relevant for technical applications like soldering \cite{BMBM1995pf,KuPr2007aci}, but also have many biophysical applications such as, e.g., intracellular processes related to the wetting properties of membranes and biomolecular condensates \cite{BLHR2017nrmcb, GKS2022n, MCZ2023nc}.

It is tempting to base theoretical descriptions of such complex dynamic scenarios, where chemical reactions couple nontrivially to interface-dominated hydrodynamic transport processes, on a simple addition of mass action chemical kinetics to existing hydrodynamic descriptions. However, there is a certain danger that such an \textit{ad hoc} approach may miss certain cross-influences of the different aspects that one has added together.
A preferential approach could be to obtain the coupled chemo-hydrodynamic dynamics from a common thermodynamic framework, namely, a gradient dynamics on an underlying energy functional as well known from the description of chemically inert systems described, e.g., by mesoscopic hydrodynamics (thin-film models) \cite{Mitl1993jcis,ThAP2016prf,HDJT2023jfm,Thie2018csa,HDGT2024l}. If chemical reactions were incorporated into such an approach, systems would still be relaxational, i.e., would approach an equilibrium. Then, specific nonequilibrium conditions could be included in a well-controlled manner. In this way, couplings that originate in thermodynamic constraints and that may significantly affect statics and dynamics are still present, even when the system is permanently driven away from equilibrium, as is often the case in biological contexts. Familiar examples of reciprocal pairs of cross-couplings include the Seebeck and Peltier effects (thermoelectricity) and the Soret and Dufour effects (thermophoresis and diffusion thermoeffect) \cite{Mill1960cr, RaSa2014ijhmt, GrootMazur1984}. These effects are captured by linear nonequilibrium thermodynamics and are pairwise related by the Onsager reciprocity relations \cite{Onsa1931prb, Onsa1931pr} as a consequence of microscopic reversibility (detailed balance). In general, chemical reactions lie outside of the framework of linear thermodynamics \cite{GrootMazur1984}. However, by invoking detailed balance one may still infer thermodynamically consistent kinetics which then for ideal systems recover mass action kinetics \cite{Marcelin1914,DeDonder1936,Zwic2022cocis}.

Returning specifically to the description of reactive wetting and reactive surfactants, we may then find chemical reactions \reviewchange{that not only depend} on the reactant concentrations but also on, e.g., reactant-substrate interactions, the proximity of the three-phase contact line or the shape of the liquid-gas interface. In the present work, we \reviewchange{show} how chemical reactions can be incorporated into a gradient dynamics model to obtain a description of reactive thin-film hydrodynamics responsible, e.g., for reactive wetting. This approach is based on an underlying (free) energy functional from which thermodynamically fully consistent kinetic equations describing reactive thin-film hydrodynamics are derived.

The structure of our work is as follows. In Section~\ref{sec:general_linear_gradient_dynamics}, we review the gradient dynamics approach to linear nonequilibrium thermodynamics. In Section~\ref{sec:chemical_reactions_as_gd}, we turn to the gradient dynamics description of chemical reactions. We start with a simple autocatalytic reaction in a spatially homogeneous system, clarify how mass action kinetics may be re-expressed in terms of a free energy functional and show that the principle of detailed balance guarantees thermodynamic consistency. We also discuss how the breaking of detailed balance, e.g.~by chemostatting, can result in persistent nonequilibrium  behavior. Section~\ref{sec:applications_reactive_wetting} then exemplifies the approach by considering two cases of reactive mesoscopic thin-film hydrodynamics. Both involve the interplay of chemical reactions and interfacial effects, e.g., reactant-induced wettability gradients or solutal Marangoni stresses. Further, we show how breaking the gradient dynamics form by external chemostats results in active behavior such as the self-propulsion of drops and the self-excitation of various drop oscillations. We conclude in Section~\ref{sec:conclusion} with a summary and discussion of possible extensions.
\section{Linear gradient dynamics and applications}\label{sec:general_linear_gradient_dynamics}
Here, we briefly review the gradient dynamics description of systems that are sufficiently close to thermodynamic equilibrium to allow for a description by linear nonequilibrium thermodynamics, i.e., the kinetic equations are linear in the variations of an underlying thermodynamic functional (the thermodynamic forces). In Section~\ref{sec:general_equations_linear_gradient_dynamics}, we introduce the general kinetic equations for \reviewchange{the dynamics of a system of $N$ scalar fields} and show their thermodynamic consistency. In Section~\ref{sec:examples_linear_gradient_dynamics} we give two corresponding examples related to reactive thin-film hydrodynamics.
\subsection{General Equations}\label{sec:general_equations_linear_gradient_dynamics}
We now consider a spatially extended system, the state of which is fully determined by $N$ scalar state variables $\vb{u}=\left(u_1, u_2, \ldots, u_N\right)^\text{T}$ which depend on the spatial coordinates $\vec{x}=(x, y, z)^\text{T}$ and time $t$. We assume that the system relaxes to thermodynamic equilibrium. Depending on the relevant constraints, different thermodynamic functionals capture this relaxation process in agreement with the second law of thermodynamics. Here, we restrict ourselves to closed, isothermal systems. In this case, the (Helmholtz) free energy functional $F\left[\vb{u}\right]$ decreases monotonically until equilibrium is reached. If the system is sufficiently close to thermodynamic equilibrium, the general time evolution equations for the fields $\vb{u}$ can be expressed as
\begin{equation}
\partial_t u_\alpha = \nabla\cdot\left[\sum_{\beta=1}^N Q^\mathrm{c}_{\alpha\beta}\nabla\frac{\delta F}{\delta u_\beta}\right]-\sum_{\beta=1}^N Q^\mathrm{nc}_{\alpha\beta} \frac{\delta F}{\delta u_\beta}, \label{eq:general_linear_gradient_dynamics}
\end{equation}
where $\partial_t$ denotes the partial derivative with respect to time $t$ and $\nabla = \left(\partial_x, \partial_y, \partial_z\right)^\text{T}$ is the spatial gradient operator. The variational derivatives $\delta F/\delta u_\beta$ express thermodynamically conjugate quantities to the fields $u_\beta$ such as the pressure or chemical potentials. In (\ref{eq:general_linear_gradient_dynamics}), these variations linearly enter the time evolution equations. Here, the first term corresponds to transport processes that conserve $\int u_\alpha\text{d}^3r $ and that are driven by spatial gradients in pressure or chemical potentials. The second term is associated with nonconserved contributions to the dynamics that are related to transitions between the individual fields and are, for instance, driven by local differences in chemical potentials. The mobility matrices $\tens{Q}^\mathrm{c} = \left(Q^\mathrm{c}_{\alpha\beta}\right)$ and $\tens{Q}^\mathrm{nc} = \left(Q^\mathrm{nc}_{\alpha\beta}\right)$ relate the energy variations to the fluxes and are positive (semi-)definite and symmetric, expressing irreversibility of the macroscopic processes and microscopic reversibility (Onsager relations \mbox{\cite{Onsa1931prb, Onsa1931pr}}). We note that the specific form (\ref{eq:general_linear_gradient_dynamics}) can be derived from Onsager's variational principle \cite{Doi2011jpcm, Thie2018csa}. Using (\ref{eq:general_linear_gradient_dynamics}) and the properties of $\tens{Q}^\mathrm{c},\tens{Q}^\mathrm{nc}$, one can show that the free energy $F$ decreases monotonically:
\begin{align}
\frac{\text{d}F}{\text{d}t} &= \int \left[\sum_{\alpha=1}^N \frac{\delta F}{\delta u_\alpha}\frac{\partial u_\alpha}{\partial t}\right]\text{d}^3r \label{eq:linear_GD_proof_line1}\\
&=\int \left[\sum_{\alpha , \beta=1}^N \frac{\delta F}{\delta \reviewchange{u}_\alpha}\nabla \cdot \left(Q^\mathrm{c}_{\alpha\beta}\nabla\frac{\delta F}{\delta \reviewchange{u}_\beta}\right)\right]\text{d}^3r - \int \left[\sum_{\alpha , \beta=1}^N \frac{\delta F}{\delta \reviewchange{u}_\alpha} Q^\mathrm{nc}_{\alpha\beta} \frac{\delta F}{\delta \reviewchange{u}_\beta}\right]\text{d}^3r \label{eq:linear_GD_proof_line2}\\
&=-\int \left[\sum_{\alpha , \beta=1}^N Q^\mathrm{c}_{\alpha\beta}\left(\nabla\frac{\delta F}{\delta \reviewchange{u}_\alpha}\right) \cdot \left(\nabla\frac{\delta F}{\delta \reviewchange{u}_\beta}\right)\right]\text{d}^3r - \int \left[\sum_{\alpha , \beta=1}^N \frac{\delta F}{\delta \reviewchange{u}_\alpha} Q^\mathrm{nc}_{\alpha\beta} \frac{\delta F}{\delta \reviewchange{u}_\beta}\right]\text{d}^3r\label{eq:linear_GD_proof_line3}\\
& \leq 0.
\end{align}
Here, integration is performed over the whole system domain. From (\ref{eq:linear_GD_proof_line1}) to (\ref{eq:linear_GD_proof_line2}) we have used the \reviewchange{dynamic equations} (\ref{eq:general_linear_gradient_dynamics}) and from (\ref{eq:linear_GD_proof_line2}) to (\ref{eq:linear_GD_proof_line3}) partial integration was used on the first term, assuming no-flux or periodic boundary conditions. The final inequality follows from the positive (semi-)definiteness of the mobility matrices. Then, if $F$ is bounded from below, it is a Lyapunov functional to the dynamics (\ref{eq:general_linear_gradient_dynamics}) which is therefore consistent with (linear) nonequilibrium thermodynamics.
\subsection{Examples}\label{sec:examples_linear_gradient_dynamics}
To model particular systems, the described gradient dynamics model (\ref{eq:general_linear_gradient_dynamics}) can be supplemented by specific choices for the energy functional and the mobility matrices. In the following, we briefly discuss a few models. For instance, the diffusion of an ideal gas is described by the diffusion equation
\begin{equation}
\partial_t n = D\Delta n,\label{eq:diffusion}
\end{equation}
where $\Delta=\partial_x^2+\partial_y^2+\partial_z^2$ is the Laplace operator, $n$ denotes the local particle density (particles per unit volume) and $D>0$ is the diffusion constant. With the entropic free energy
\begin{equation}
F = k_b T \int n\left[\ln(n/n_0)-1\right]\text{d}^3r,
\end{equation}
where $k_b$ is the Boltzmann constant, $T$ is the temperature, $n_0$ is some reference density, and with the diffusive mobility
\begin{equation}
M(n) = \frac{D n}{k_b T},
\end{equation}
the diffusion equation (\ref{eq:diffusion}) can be brought into the gradient dynamics form
\begin{equation}
\partial_t n = \nabla\cdot\left[M(n)\nabla\frac{\delta F}{\delta n}\right].
\end{equation}
Note that here $\delta F/\delta n$ corresponds to a chemical potential.

Mesoscopic thin-film equations \cite{Thie2018csa} represent another class of such models. In the simplest case, the evolution of the local film height $h$ of a thin liquid film on a homogeneous rigid solid substrate is described by \cite{OrDB1997rmp,Thie2007chapter,BEIM2009rmp}
\begin{equation}
\partial_t h = -\nabla\cdot\left[Q(h)\nabla\left(\gamma\Delta h +\Pi(h)\right)\right]\label{eq:simple_TF_equation}
\end{equation}
with $\nabla = (\partial_x, \partial_y)^\text{T}$ and $\Delta = \partial_x^2+\partial_y^2$. Here, $\gamma$ is a constant surface tension and the term $-\gamma\Delta h$ is the Laplace pressure, while $\Pi(h)$ is a Derjaguin (disjoining) pressure that encodes wettability on the mesoscopic scale \cite{Genn1985rmp,StVe2009jpm,Thie2010jpcm}. The mobility factor $Q(h)\geq 0$ typically is a polynomial in $h$ and depends on the particular model, e.g., the degree of slip at the substrate \cite{MiDa1994jfm,OrDB1997rmp,MuWW2005jem}. We introduce the free energy functional
\begin{equation}
F = \int\left[f(h)+\frac{\gamma}{2}\vert\nabla h\vert ^2\right]\text{d}^2 r,
\end{equation} 
where the Derjaguin pressure is $\Pi(h)=-f'(h)$ with the \reviewchange{prime} indicating the univariate derivative. We can then write (\ref{eq:simple_TF_equation}) as \cite{Mitl1993jcis,Thie2010jpcm}
\begin{equation}
\partial_t h = \nabla\cdot\left[Q(h)\nabla\frac{\delta F}{\delta h}\right]. \label{eq:TF_as_GD}
\end{equation}
From (\ref{eq:TF_as_GD}), we can see that $\delta F/\delta h$ corresponds to a pressure (or its negative). Besides the two shown simple examples, other well-known examples include the Cahn-Hilliard equation \cite{CaHi1958jcp, Bray1994ap}, various two- and three-field thin-film models, e.g.,  for two-layer films, films of binary mixtures, surfactant-covered drops and drops on soft and adaptive substrates \cite{PBMT2005jcp,ThAP2012pf, ThTL2013prl, HJKP2015jem,ThAP2016prf, ThHa2020epjt, HeST2021sm,HDGT2024l} as well as other soft matter models \cite{ArRa2004jpag,Doi2011jpcm}.
\section{Chemical reactions as nonlinear gradient dynamics}\label{sec:chemical_reactions_as_gd}
After having treated the gradient dynamics form of systems that are captured by linear nonequilibrium thermodynamics in Section~\ref{sec:general_linear_gradient_dynamics}, we now turn to systems allowing for chemical reactions. As we will see, reactive processes that follow mass action type kinetics lie outside of the framework of linear nonequilibrium thermodynamics, i.e., the corresponding reactive fluxes between chemical species are nonlinear in the variations of the underlying thermodynamic functional.\footnote{We remark that with somewhat `unnatural' mobilities that contain variations, chemical reactions can be treated by means of linear nonequilibrium thermodynamics \cite{GrOe1997pre, OeGr1997pre, Miel2011n} while the resulting fluxes are still nonlinear in the variations.} To motivate this, we first treat ideal mixtures in Section~\ref{sec:ideal_mixtures_mass_action_as_gd} and show that the corresponding mass action type kinetics can be rewritten employing rates that are nonlinear in the (free) energy variations. We then show that the general dynamic equations indeed represent a gradient dynamics when assuming detailed balanced kinetics. In Section~\ref{sec:non_ideal_cross_couplings}, we discuss possible deviations from ideal mass action type kinetics and touch on some of the reactive cross-couplings that follow.
\subsection{Mass action type kinetics as gradient dynamics}\label{sec:ideal_mixtures_mass_action_as_gd}
To motivate that chemical processes are insufficiently treated by linear nonequilibrium thermodynamics, we consider a closed, isothermal box of volume $V$ at temperature $T$ containing two well-mixed ideal gases, namely species~$N_1$ and $N_2$. For this setup, the (Helmholtz) free energy appropriately describes the approach to thermodynamic equilibrium. To be specific, we assume that the autocatalytic conversion reaction
\begin{equation}
2N_1 + N_2 \rightleftharpoons 3N_1 \label{eq:autocatalysis_reaction}
\end{equation}
takes place between the two chemical species. For mixtures of ideal gases, chemical reactions are appropriately described by mass action kinetics \cite{GrootMazur1984}. The time evolution of the particle densities $n_1$ and $n_2$ of the two gases (particles per unit volume) inside the box is then given by
\begin{align}
\frac{\text{d}n_1}{\text{d}t} &= r_\mathrm{f} n_1^2 n_2 - r_\mathrm{b} n_1^3, \label{eq:time_ev_well_mixed_reactor_n1}\\
\frac{\text{d}n_2}{\text{d}t} &= -r_\mathrm{f} n_1^2 n_2 + r_\mathrm{b} n_1^3, \label{eq:time_ev_well_mixed_reactor_n2}
\end{align}
where $r_\mathrm{f}, r_\mathrm{b} >0$ denote the reaction rates in the forward and backward direction. Thereby, the reaction $2N_1 + N_2 \rightarrow 3N_1$ is defined as the forward reaction. Note that $\text{d}(n_1+n_2)/\text{d}t=0$ and the total particle \reviewchange{number} $N = (n_1+n_2)V$ in the container is conserved. We consider the associated chemical potentials of ideal gases \cite{GrootMazur1984}
\begin{equation}
\mu_\alpha = k_b T \ln\left(n_\alpha/n_{0,\alpha}\right)+\zeta_\alpha(T), \label{eq:ideal_chemical_potential}
\end{equation}
where $n_{0,\alpha}$ are reference densities and $\zeta_\alpha(T)$ are general dependencies on the temperature $T$. Note that $\zeta_\alpha$ represents the chemical potential of species $\alpha$ at the reference density $n_{0,\alpha}$. For simplicity, we set $\zeta_\alpha = 0$ in the following. Using Eq.~(\ref{eq:ideal_chemical_potential}), we rewrite Eqs.~(\ref{eq:time_ev_well_mixed_reactor_n1}) and (\ref{eq:time_ev_well_mixed_reactor_n2}) as
\begin{align}
\frac{\text{d}n_1}{\text{d}t} &= r_\mathrm{f} n^2_{0,1}n_{0,2} e^{\frac{2\mu_1+\mu_2}{k_b T}} - r_\mathrm{b} n^3_{0,1} e^{\frac{3\mu_1}{k_b T}}, \label{eq:time_ev_well_mixed_reactor_n1_mu}\\
\frac{\text{d}n_2}{\text{d}t} &= -r_\mathrm{f} n^2_{0,1}n_{0,2} e^{\frac{2\mu_1+\mu_2}{k_b T}} + r_\mathrm{b} n^3_{0,1} e^{\frac{3\mu_1}{k_b T}}. \label{eq:time_ev_well_mixed_reactor_n2_mu}
\end{align}
We know that the gas mixture must approach thermodynamic equilibrium. For chemical reactions thermodynamic equilibrium is equivalent to the condition that the affinity $A$ vanishes \cite{GrootMazur1984}. The affinity corresponds to the difference in total chemical energy between the reactants and the products of a reaction and represents the thermodynamic force that drives the reaction.\footnote{Note that the affinity $A_j$ can be related to the reaction coordinate $\xi_j$ which represents the `degree of advancement' of reaction $j$ (see also \cite{GrootMazur1984}). In a well-mixed reactor it is related to the particle number by $N_\alpha = N_{0,\alpha}+\sum_j (\nu^b_{\alpha j}-\nu^f_{\alpha j})\xi_j$ where $N_{0,\alpha}$ are the \reviewchange{initial particle numbers} or equivalently $\partial\xi_j/\partial t = J_j$. Differentiating, e.g., the free energy $F$ with respect to $\xi_j$ we obtain $\left(\partial F/\partial\xi_j\right)_{T,V, \xi_{i\neq j}}=\sum_\alpha\left(\partial F / \partial N_\alpha \right)_{T,V, N_{\beta\neq\alpha}}\partial N_\alpha / \partial\xi_j = \sum_\alpha (\nu^b_{\alpha j}-\nu^f_{\alpha j})\mu_\alpha= A_j$, that is, the affinity is the derivative of the relevant thermodynamic potential with respect to the reaction coordinate \cite{DeDonder1936}.}
Specifically for reaction (\ref{eq:autocatalysis_reaction}) the affinity is given by $A=3\mu_1-(2\mu_1+\mu_2) = \mu_1-\mu_2$. Setting the right hand side of (\ref{eq:time_ev_well_mixed_reactor_n1_mu})-(\ref{eq:time_ev_well_mixed_reactor_n2_mu}) to zero for $A=0$, one finds that this condition only holds if
$r = r_\mathrm{f} n^2_{0,1} n_{0,2} = r_\mathrm{b} n^3_{0,1}$ is the \reviewchange{rate} for both the forward and the backward reaction. Only then
(\ref{eq:time_ev_well_mixed_reactor_n1_mu})-(\ref{eq:time_ev_well_mixed_reactor_n2_mu}) can be reformulated as
\begin{align}
\frac{\text{d}n_1}{\text{d}t} &= r\left(e^{\frac{2\mu_1+\mu_2}{k_b T}} - e^{\frac{3\mu_1}{k_b T}}\right), \label{eq:time_ev_well_mixed_reactor_n1_DB}\\
\frac{\text{d}n_2}{\text{d}t} &= -r\left(e^{\frac{2\mu_1+\mu_2}{k_b T}} - e^{\frac{3\mu_1}{k_b T}}\right), \label{eq:time_ev_well_mixed_reactor_n2_DB}
\end{align}
with a common \reviewchange{rate (function) $r$}.\footnote{Note that we have restricted ourselves to mixtures of ideal gases in the derivation of (\ref{eq:time_ev_well_mixed_reactor_n1_DB})-(\ref{eq:time_ev_well_mixed_reactor_n2_DB}). Analo\-gously, one could consider a reacting ideal solution (or any other ideal system), for which mass action kinetics are also commonly assumed. The chemical potential of, e.g., component $1$ in an ideal solution would read $\mu_1 = k_b T\ln(\frac{n_1}{n})+\zeta_1(T)$ with the total particle density $n_1+n_2 = n$ and where the first term corresponds to the mixing entropy of $N_1$. If one starts with mass action kinetics formulated in terms of particle densities as in (\ref{eq:time_ev_well_mixed_reactor_n1})-(\ref{eq:time_ev_well_mixed_reactor_n2}), then the equilibrium constant $r_\mathrm{f}/r_\mathrm{b}$ generally becomes a function of $T$ \textit{and} $n$ \cite{KoudriavtsevJamesonLinert2001, GrootMazur1984}. However, this does not change the following general results and the corresponding mass action kinetics can still be derived from Eqs. (\ref{eq:time_ev_well_mixed_reactor_n1_DB})-(\ref{eq:time_ev_well_mixed_reactor_n2_DB}) or (\ref{eq:time_ev_well_mixed_reactor_n1_func_deriv})-(\ref{eq:time_ev_well_mixed_reactor_n2_func_deriv}).}  Writing the reactive flux as
\begin{equation}
J = J^f-J^b = r\left(e^{\frac{2\mu_1+\mu_2}{k_b T}} - e^{\frac{3\mu_1}{k_b T}}\right), \label{eq:reactive_flux_autocatalysis}
\end{equation}
at thermodynamic equilibrium we have equal fluxes in both reactive directions, i.e.,
\begin{equation}
J^f_{\text{eq}} = J^b_{\text{eq}}. \label{eq:detailed_balance_condition}
\end{equation}
Condition (\ref{eq:detailed_balance_condition}) is the principle of detailed balance, which follows from microscopic reversibility \cite{Gorb2014rp}, and here corresponds to the common \reviewchange{rate function} $r$ for \textit{both reactive directions}. While (\ref{eq:detailed_balance_condition}) seems rather trivial, as it is simply the steady state condition for one reaction, in systems with multiple reactions one may also find steady states for which the reactive fluxes do not vanish \textit{separately}, \reviewchange{as discussed} by Wegscheider \citep{Wegs1902zpc}. These states need to be distinguished from genuine thermodynamic equilibrium given by detailed balance (\ref{eq:detailed_balance_condition}). We note that by introducing the free energy functional
\begin{equation}
F = \int_V\left(k_b T n_1\left[\ln(n_1/n_{0,1})-1\right]+k_b T n_2\left[\ln(n_2/n_{0,2})-1\right]\right)\text{d}V,
\end{equation}
we can express the chemical potentials (\ref{eq:ideal_chemical_potential}) as $\mu_\alpha = \frac{\delta F}{\delta n_\alpha}$ and therefore obtain as time evolution equations
\begin{align}
\frac{\text{d}n_1}{\text{d}t} &= r\left[\exp\left(\frac{2\frac{\delta F}{\delta n_1}+\frac{\delta F}{\delta n_2}}{k_b T}\right) - \exp\left(3\frac{\frac{\delta F}{\delta n_1}}{k_b T}\right)\right], \label{eq:time_ev_well_mixed_reactor_n1_func_deriv}\\
\frac{\text{d}n_2}{\text{d}t} &= -r\left[\exp\left(\frac{2\frac{\delta F}{\delta n_1}+\frac{\delta F}{\delta n_2}}{k_b T}\right) - \exp\left(3\frac{\frac{\delta F}{\delta n_1}}{k_b T}\right)\right]. \label{eq:time_ev_well_mixed_reactor_n2_func_deriv}
\end{align}
The more general form of (\ref{eq:time_ev_well_mixed_reactor_n1_func_deriv}) and (\ref{eq:time_ev_well_mixed_reactor_n2_func_deriv}) dates back to Marcelin \cite{Marcelin1914} and De Donder \cite{DeDonder1936, Van1958jcp} and suggests applicability to nonideal systems and to more complicated energy functionals as well as \reviewchange{heterogeneous} spatially extended systems. While we have motivated the form (\ref{eq:time_ev_well_mixed_reactor_n1_func_deriv})-(\ref{eq:time_ev_well_mixed_reactor_n2_func_deriv}) starting from simple mass action kinetics, we note that the same expression can be derived by considering chemical reactions as a diffusion process in the (continuous) configuration space of the reactive complex so that this form indeed remains valid for nonideal systems  \cite{PaPR1997pa, HaTB1990rmp, Peters2017}. \reviewchange{Note that in general, the \reviewchange{rate function} $r$ need not be constant and may depend on,~e.g., the local concentrations.} We now show that detailed balanced mass action type kinetics such as Eqs.~(\ref{eq:time_ev_well_mixed_reactor_n1_func_deriv})-(\ref{eq:time_ev_well_mixed_reactor_n2_func_deriv}) is a gradient dynamics, i.e., that $F$ monotonically decreases. 
To this end, we treat a more general scenario and consider a \reviewchange{heterogeneous} spatially extended system comprising the components $N_1, \ldots, N_Q$ with the particle densities $n_1, \ldots, n_Q$ in a closed, thermostatted box of volume $V$. \reviewchange{For clarity of notation, we exclusively use Greek script for field variable indexing and Latin script for reaction indexing. Between the $Q$ components, $R$ chemical reactions may occur. Each reaction $j$ is characterized by} its stoichiometric coefficients $\nu_{\beta j}^{f},\nu_{\beta j}^{b}\geq 0$ for each component $\beta$ in the forward and backward directions, respectively. Each reaction $j$ can then be summarized as 
\begin{equation}
\sum_{\beta=1}^{Q} \nu^f_{\beta j} N_\beta \leftrightharpoons \sum_{\beta=1}^{Q} \nu^b_{\beta j} N_\beta. \label{eq:general_reaction_scheme}
\end{equation}
Note that because chemical reactions conserve total mass, the stoichiometric coefficients must obey the condition
\begin{equation}
\sum_{\beta=1}^Q m_\beta\nu^f_{\beta j} = \sum_{\beta=1}^Q m_\beta\nu^b_{\beta j},\label{eq:chemical_reaction_mass_conservation}
\end{equation}
where $m_\beta$ is the mass per particle for component $\beta$. Eq.~(\ref{eq:chemical_reaction_mass_conservation}) states that the total mass of reactants and products must be identical. For instance, applying (\ref{eq:chemical_reaction_mass_conservation}) to the autocatalysis reaction (\ref{eq:autocatalysis_reaction}) implies that the molecular masses must be equal. In addition to conservation of the total mass $M = \int_V\left[m_1 n_1+m_2 n_2\right]\text{d}V$ the total particle \reviewchange{number} $N = \int_V\left[n_1+n_2\right]\text{d}V$ is then also conserved, as already observed from Eqs.~(\ref{eq:time_ev_well_mixed_reactor_n1})-(\ref{eq:time_ev_well_mixed_reactor_n2}). One should always take care that the particular choices of molecular masses (which may, e.g., explicitly appear in some transport coefficients) and of stoichiometric coefficients do not contradict mass conservation. In analogy to (\ref{eq:reactive_flux_autocatalysis}), we associate with each reaction $j$, specified by (\ref{eq:general_reaction_scheme}), a detailed balanced mass action type kinetics
\begin{equation}
J_{j}=J_j^f-J_j^b = r_{j}\left[\exp\left(\frac{\sum_{\beta=1}^Q\nu^f_{\beta j}\frac{\delta F}{\delta n_\beta}}{k_b T}\right)-\exp\left(\frac{\sum_{\beta=1}^Q\nu^b_{\beta j}\frac{\delta F}{\delta n_\beta}}{k_b T}\right)\right],\label{eq:general_db_mass_action_kinetics}
\end{equation}
where $r_{j}>0$ is again the common \reviewchange{rate function} for both reaction directions and $F$ is the free energy (or any other appropriate thermodynamic functional). Note that (\ref{eq:general_db_mass_action_kinetics}) can be expanded close to chemical equilibrium ($A=\sum_{\beta=1}^Q(\nu^b_{\beta j}-\nu^f_{\beta j})\frac{\delta F}{\delta n_\beta} \ll k_b T$) to obtain a flux that is linear in the affinity $A$ \cite{GrootMazur1984}. To demonstrate applicability to spatially \reviewchange{heterogeneous} extended systems, we additionally assume diffusive transport such that the total time evolution of the particle densities is given by
\begin{equation}
\frac{\partial n_\alpha}{\partial t} = \sum_{\beta=1}^Q \nabla \cdot \left(L_{\alpha\beta}\nabla\frac{\delta F}{\delta n_\beta}\right)-\sum_{j=1}^R (\nu^f_{\alpha j}-\nu^b_{\alpha j}) J_j, \label{eq:RD_gradient dynamics}
\end{equation}
where the diffusive coefficients $L_{\alpha\beta}$ form a symmetric, positive (semi)-definite matrix and $\nabla = (\frac{\partial}{\partial x}, \frac{\partial}{\partial y}, \frac{\partial}{\partial z})^\text{T}$. Note that (\ref{eq:RD_gradient dynamics}) conserves the total mass $M = \int_V\left[\sum_{\beta=1}^Q m_\beta n_\beta\right]\text{d}V$ due to (\ref{eq:chemical_reaction_mass_conservation}). 
In general, the total number of conserved quantities is determined as $Q-S$, where $S$ is the dimension of the stoichiometric subspace $\mathcal{S}$ that is spanned by the vectors $\boldsymbol{\nu}_j=(\nu_{\alpha j})=(\nu^{f}_{\alpha j}-\nu^{b}_{\alpha j})$. Conserved quantities are then given as linear combinations $\sum_\alpha c_\alpha n_\alpha$, where $\vb{c}=(c_\alpha)$ are vectors in the orthogonal complement of $\mathcal{S}$, i.e., $\vb{c}\cdot \boldsymbol{\nu}_j=0$ for all $j$ \cite{Feinberg2019}. Since chemical reactions conserve mass, this implies that particular behavior described for reaction-diffusion systems with conservation laws \cite{JoBa2005pb,IsOM2007pre,MoOg2010n,AlBa2010pb,KuMo2015pre,BDGY2017nc,HaFr2018np,BrHF2020prx,BWHY2021prl} is most likely a better representation of generic behavior than the more frequently investigated fully open reaction-diffusion systems without any conservation law.

We next show that (\ref{eq:RD_gradient dynamics}) constitutes a gradient dynamics. For the free energy $F$ we have
\begin{align}
\frac{\text{d}F}{\text{d}t} &= \int_V \left[\sum_{\beta=1}^Q \frac{\delta F}{\delta n_\beta}\frac{\partial n_\beta}{\partial t}\right]\text{d}^3 r \label{eq:GD_proof_line1}\\
&=\int_V \left[\sum_{\alpha , \beta=1}^Q \frac{\delta F}{\delta n_\beta}\nabla \cdot \left(L_{\alpha\beta}\nabla\frac{\delta F}{\delta n_\alpha}\right)\right]\text{d}V - \sum_{j=1}^R \int_V \left[\sum_{\beta=1}^Q \frac{\delta F}{\delta n_\beta}(\nu^f_{\beta j}-\nu^b_{\beta j}) J_j \right]\text{d}^3 r \label{eq:GD_proof_line2}\\
&=-\int_V \left[\sum_{\alpha , \beta=1}^Q L_{\alpha\beta} \left(\nabla \frac{\delta F}{\delta n_\beta}\right)\cdot \left(\nabla\frac{\delta F}{\delta n_\alpha}\right)\right]\text{d}V - k_b T \sum_{j=1}^R \int_V \ln\left(J^f_j/J^b_j\right) J_j\:\text{d}^3 r \label{eq:GD_proof_line3}\\
&=-\int_V \left[\sum_{\alpha , \beta=1}^Q L_{\alpha\beta} \left(\nabla \frac{\delta F}{\delta n_\beta}\right)\cdot \left(\nabla\frac{\delta F}{\delta n_\alpha}\right)\right]\text{d}V - k_b T \sum_{j=1}^R \int_V \ln\left(J^f_j/J^b_j\right) (J_j^f-J_j^b)\:\text{d}^3 r \label{eq:GD_proof_line4}\\
& \leq 0.\label{eq:GD_proof_line5}
\end{align}
From (\ref{eq:GD_proof_line1}) to (\ref{eq:GD_proof_line2}) we have used the time evolution equations (\ref{eq:RD_gradient dynamics}). In (\ref{eq:GD_proof_line3}), partial integration was performed on the first term assuming, e.g., no-flux boundary conditions. The second term was re-expressed in (\ref{eq:GD_proof_line3}) and (\ref{eq:GD_proof_line4}) using Eq.~(\ref{eq:general_db_mass_action_kinetics}). We stress that the trans\-for\-mation from (\ref{eq:GD_proof_line2}) to (\ref{eq:GD_proof_line3}) requires a common \reviewchange{rate function} $r_j$ for each reaction $j$ and is therefore only possible for detailed balanced kinetics. The final inequality follows from the positive definiteness of the transport matrix $\tens{L}= (L_{\alpha\beta})$ and from the inequality $\left[f(b)-f(a)\right]\left[b-a\right]\geq 0$ for any monotonic $f$ if $b\geq a$. We thus conclude that any bounded $F$ is a Lyapunov functional to the dynamics (\ref{eq:RD_gradient dynamics}) with its minimum corresponding to thermodynamic equilibrium\footnote{For an analogous proof for the entropy production, see for example~\cite{Gaspard2022}.}. There exist several ways to go beyond the described relaxational dynamics, i.e., to obtain sustained out-of-equilibrium dynamics (e.g., oscillatory dynamics), as found in many active systems. A common strategy consists of chemostatting one or several of the reactive components, i.e., keeping them at a constant chemical potential \cite{KiZw2021jotrsi, CoGA2022prl, Zwic2022cocis}. This breaks detailed balance (\ref{eq:general_db_mass_action_kinetics}) since, by rearranging the indices of the species, we can write the fluxes as 
\begin{align}
J_{j}&= r_{j}\left[\exp\left(\frac{\sum_{\beta=1}^{Q'}\nu^f_{\beta j}\frac{\delta F}{\delta n_\beta}+\sum_{\reviewchange{\beta=Q'+1}}^{\reviewchange{Q}}\nu^f_{\beta j}\mu_{\beta ,0}}{k_b T}\right)-\exp\left(\frac{\sum_{\beta=1}^{Q'}\nu^b_{\beta j}\frac{\delta F}{\delta n_\beta}+\sum_{\reviewchange{\beta=Q'+1}}^{\reviewchange{Q}}\nu^b_{\beta j}\mu_{\beta ,0}}{k_b T}\right)\right]\\ \label{eq:general_chemostatting_fluxes}
&=\tilde{r}^f_{j}\exp\left(\frac{\sum_{\beta=1}^{Q'}\nu^f_{\beta j}\frac{\delta F}{\delta n_\beta}}{k_b T}\right)-\tilde{r}^b_{j}\exp\left(\frac{\sum_{\beta=1}^{Q'}\nu^b_{\beta j}\frac{\delta F}{\delta n_\beta}}{k_b T}\right),
\end{align}
\reviewchange{where $Q'$ is the number of nonchemostatted species and $Q-Q'$ is the number of chemostatted species that are kept at the respective chemical potentials $\mu_{\beta,0}$}. The constant chemical potentials can be absorbed into the effective rates
\begin{align}
\tilde{r}^f_{j} &= r_j\exp\left(\frac{\sum_{\reviewchange{\beta=Q'+1}}^{\reviewchange{Q}}\nu^f_{\beta j}\mu_{\beta ,0}}{k_b T}\right),\\
\tilde{r}^b_{j} &= r_j\exp\left(\frac{\sum_{\reviewchange{\beta=Q'+1}}^{\reviewchange{Q}}\nu^b_{\beta j}\mu_{\beta ,0}}{k_b T}\right),
\end{align}
which are generally distinct such that the proof (\ref{eq:GD_proof_line1})-(\ref{eq:GD_proof_line5}) breaks down. Alternatively, one may directly break detailed balance by introducing different rates $r^f_j, r^b_j$ in the forward and backward directions or by simply assuming irreversible reactions ($r^b_j = 0$ for some $j$). This may be appropriate, e.g., when treating active protein and enzymatic reactions \cite{DGMF2023prl, BWWF2022nrp, GoLA2005prl} or in open systems if the reaction products are immediately removed. A combination of both strategies can be found in the construction of the Brusselator model \cite{PrLe1968jcp, Nicolis1999}, where the presence of irreversible reactions and assumed constant concentrations of certain reactants lead to chemical oscillations. Finally, we note that breaking the principle of detailed balance does not necessarily result in models that allow for `active dynamics'. Even for (well-mixed) open systems, there exist wide classes of mass action reaction networks with  `complex balance' that guarantees the existence of a Lyapunov-function and a unique equilibrium point which, however, does not coincide with thermodynamic equilibrium in general \cite{HoJa1972arma, Feinberg2019, RaEs2016prx}. 
\subsection{Reactive cross-couplings in nonideal systems}
\label{sec:non_ideal_cross_couplings} 
Having shown that fluxes of the form (\ref{eq:general_db_mass_action_kinetics}) result in a proper gradient dynamics, they may now be used to incorporate chemical reactions into models of systems with more complicated free energies and transport processes possibly resulting in chemo-mechanical coupling. For nonideal systems, Eq.~(\ref{eq:general_db_mass_action_kinetics}) then yields mass action type kinetics with reactive fluxes that contain additional factors from nonideal or even mechanical contributions to the chemical potentials. A common example comprises the interaction between chemical reactions and phase separation \cite{KiZw2021jotrsi, ZiKZ2023prl, Zwic2022cocis}. There, the free energy is amended, e.g., by adding Flory-Huggins-type interactions between the different species \cite{Flor1942jcp, Hugg1941jcp} as well as interface energies. In consequence, the reactive fluxes additionally depend on the presence or absence of interfaces and on the interaction parameters. In the context of reactive wetting, reactions may be influenced by the proximity of the solid-liquid interface due to reactant-substrate interactions which may become crucial in the contact-line region. Examples include effects of reactant-dependent wettability \cite{BaBM1994n,DoOn1995prl,BrGe1995crassi,Genn1998pa,LeLa2000jacs,LeKL2002pre,ThJB2004prl,SKYN2005pre,JoBT2005epje,SuMY2006ptps,Kita2006ptps,YBJZ2012sm,Arsc2016l} or  of substrate-mediated condensation \cite{SpCR1994el,KGFC2010l,WTGK2015mmnp}. Also, for reactions of surface-active species (surfactants), the reactive fluxes can depend on the shape of the liquid-gas interface, leading to an effective chemo-mechanical coupling. This may be relevant to systems studied, e.g., in \cite{ReRV1994pf,NiAS2006pf,SuYo2014epjt,NaKI2016csaea}.
\section{Applications to reactive thin-film hydrodynamics}\label{sec:applications_reactive_wetting}
Next, we illustrate the gradient dynamics approach from Section~\ref{sec:ideal_mixtures_mass_action_as_gd} for two distinct systems involving wetting phenomena and chemical reactions that couple to interfacial physics. For simplicity, we restrict ourselves to shallow drops and thin films of partially wetting liquid on a flat, solid substrate. \reviewchange{In particular, we consider mesoscopic models that incorporate liquid-substrate interactions as a film height-dependent free energy contribution (wetting energy) which vanishes for large film heights. The final model equations then correspond to a leading-order long-wave expansion \cite{OrDB1997rmp, CrMa2009rmp} of the respective mesoscopic Stokes problem and are strictly valid only for shallow drops of heights in the nanometer to micrometer range, while qualitative agreement is also expected for larger slopes. However, we also mention that the gradient dynamics approach can naturally lead to improved thin-film models that go beyond leading order, although not in an asymptotically rigorous sense \cite{BoTH2018jfm,Thie2018csa}. This could,~e.g., be utilized in the second considered model (Section~\ref{sec:drops_with_autocatalysis}).}

In Section~\ref{sec:drops_with_wettability_gradient}, we treat drops of a suspension (or solution) where the suspended particles \reviewchange{(or molecules)} within the drop may adsorb onto the substrate and particles on the substrate may desorb into the drop. The adsorbed particles in turn modify the wettability of the substrate, leading to sustained running drops if the system is driven by chemostatting. In Section~\ref{sec:drops_with_autocatalysis}, we treat drops of a simple liquid covered by two species of insoluble surfactants which are continuously converted into each other by an autocatalytic reaction. Additionally, the drop surface is in contact with a chemostatted `bath' that acts as an external supply of surfactant. Marangoni fluxes then couple the reactive surface dynamics to the liquid bulk. As the chemostat-induced driving force is varied, we find, inter alia, Turing patterns of surfactant distributions, swaying and breathing modes of self-excited drop oscillation as well as self-propelled drops. In both cases, the gradient dynamics approach allows for thermodynamically consistent modeling, including cross-couplings and nonideal energetic contributions. 
\subsection{Reactive wetting -- Drops driven by chemically sustained wettability gradients}\label{sec:drops_with_wettability_gradient}
\begin{figure}[tbh]
	\centering
	\includegraphics[width=0.8\textwidth]{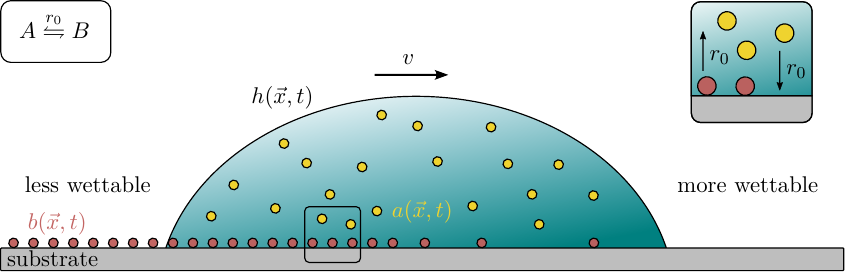}
\caption{Sketch of the considered geometry. A shallow drop of partially wetting suspension (or solution) is situated on a flat solid substate. The  particles (or the solute) may adsorb from the drop onto the substrate or desorb from the substrate into the drop, thus altering the wettability. The \reviewchange{constant rate} of this reaction is given by $r_0$ in both directions. Due to the (developing) wettability gradient, the drop moves with the velocity $v$. The local film height of the drop is $h(\vec{x}, t)$, the $z$-averaged local density of suspended particles per unit liquid volume is $a(\vec{x}, t)$ and the local density of adsorbed particles per unit substrate area is $b(\vec{x}, t)$}
\label{fig:sketch_drop_mixture_substrate_reaction}
\end{figure}

As a first example, we consider drops of a partially wetting suspension (or solution) that are situated on a flat, solid substrate (Fig.~\ref{fig:sketch_drop_mixture_substrate_reaction}). The suspended particles \reviewchange{(or molecules)} may adsorb from the drop onto the substrate and vice versa, thereby altering the wettability of the substrate. Here, we treat the case of an adsorbate that renders the substrate less wettable. Similar systems have been studied experimentally and theoretically in \cite{BaBM1994n,DoOn1995prl,BrGe1995crassi,Genn1998pa,LeLa2000jacs,LeKL2002pre,ThJB2004prl,SKYN2005pre,JoBT2005epje,SuMY2006ptps,Kita2006ptps,YBJZ2012sm,Arsc2016l} where it has been shown that the induced wettability gradient leads to self-propelled drop motion. We assume that the occuring adsorption-desorption process is appropriately described by the conversion reaction
\begin{equation}
A \stackrel{r_0}{\leftrightharpoons} B,\label{eq:adsorption_desorption_reaction}
\end{equation}
where $A$ and $B$ denote particles that are either suspended in the liquid or adsorbed onto the substrate. The corresponding particle densities per unit volume of liquid and per unit substrate area are given by $a(\vec{x}, t)$ and $b(\vec{x}, t)$, respectively. Note that for a meaningful representation in the thin-film geometry, $a(\vec{x}, t)$ must correspond to a $z$-averaged quantity. The reaction itself obeys the principle of detailed balance with the \reviewchange{constant rate} $r_0$. The local film height of the drop is $h(\vec{x}, t)$.
\noindent
With this system, we associate the free energy 
\begin{equation}
F = \int_\Omega\left[f(h, b)+\frac{\gamma}{2}\vert\nabla h\vert^2+hg_a(a)+g_b(b)+h\frac{\sigma_a}{2}\vert\nabla a\vert^2+\frac{\sigma_b}{2}\vert\nabla b\vert^2\right]\text{d}^2 r  \label{eq:F_drop_mixture_substrate_interaction}
\end{equation}
with integration over the system domain $\Omega$. In (\ref{eq:F_drop_mixture_substrate_interaction}), $f$ denotes the adsorbate-dependent wetting energy, $\gamma$ denotes the constant surface tension of the liquid,  $g_{a}$ and  $g_{b}$ are free energy contributions of $A$ and $B$ per unit liquid volume and unit substrate area, respectively, and $\sigma_{a}$, $\sigma_{b}$ denote interfacial stiffnesses that penalize gradients in $a$ and $b$. First, we compute the liquid pressure and the chemical potentials as variational derivates of (\ref{eq:F_drop_mixture_substrate_interaction}). To this end, we observe that changes in the local particle \reviewchange{number} of $A$ do not \textit{exclusively} correspond to changes in $a(\vec{x}, t)$ in the thin-film geometry. This can be seen by considering the particle \reviewchange{number} of $A$ in a liquid column of infinitesimal base area $\text{d}\Omega$, which is given by
\begin{equation}
\text{d}A = a h \cdot\text{d}\Omega = \hat{a}\cdot\text{d}\Omega\label{eq:particle_count_liquid_column_mixture}
\end{equation}
with $\hat{a} = a h$. Because the substrate area is not subject to any dynamics ($\text{d}\Omega$ \reviewchange{is constant}), Eq.~(\ref{eq:particle_count_liquid_column_mixture}) expresses a one-to-one correspondence between the particle \reviewchange{number} of species~$A$ and the field $\hat{a}$, i.e., all changes in $\text{d}A$ translate to changes in $\hat{a}$ and vice versa. Therefore, to compute the correct liquid pressure and chemical potentials, the free energy must be varied with respect to $h$, $b$ and $\hat{a}$ instead of $a$, as these fields directly relate to respective changes in the local liquid volume and the local particle \reviewchange{numbers} of $B$ and $A$. The variations are then
\begin{align}
p = \frac{\delta F}{\delta h} &= \partial_h f - \gamma\Delta h + g_a - ag_a' + \frac{\sigma_a}{2}\vert \nabla a\vert^2 + \sigma_a\frac{a}{h}\nabla\cdot\left(h\nabla a\right), \label{eq:mixture_variation_h}\\
\mu_a = \frac{\delta F}{\delta \hat{a}} &= g_a' -\frac{\sigma_a}{h}\nabla\cdot\left(h\nabla a\right),\label{eq:mixture_vartiation_a_bar}\\
\mu_b = \frac{\delta F}{\delta b} &= g_b'+\partial_b f -\sigma_b \Delta b,\label{eq:mixture_vartiation_b}
\end{align}
where $g_{a}'$ and $g_b'$ are the univariate derivates of $g_a$ and $g_b$, respectively. The \reviewchange{dynamic equations} then take on the gradient dynamics form
\begin{align}
\partial_t h &= \nabla\cdot\left[Q_{hh}\nabla\frac{\delta F}{\delta h}+Q_{ha}\nabla\frac{\delta F}{\delta \hat{a}}+Q_{hb}\nabla\frac{\delta F}{\delta b}\right]\label{eq:GD_mixture_h},\\
\partial_t \hat{a} &= \nabla\cdot\left[Q_{ah}\nabla\frac{\delta F}{\delta h}+Q_{aa}\nabla\frac{\delta F}{\delta \hat{a}}+Q_{ab}\nabla\frac{\delta F}{\delta b}\right]+J_r \label{eq:GD_mixture_a_bar},\\
\partial_t b &= \nabla\cdot\left[Q_{bh}\nabla\frac{\delta F}{\delta h}+Q_{ba}\nabla\frac{\delta F}{\delta \hat{a}}+Q_{bb}\nabla\frac{\delta F}{\delta b}\right]-J_r \label{eq:GD_mixture_b},
\end{align}
where the symmetric, positive definite transport matrix $\tens{Q}$ is
\begin{equation}
\tens{Q} = \left(\begin{array}{ccc}
Q_{hh} & Q_{ha} & Q_{hb} \\ 
Q_{ah} & Q_{aa} & Q_{ab} \\ 
Q_{bh} & Q_{ba} & Q_{bb}
\end{array} \right) = \left(\begin{array}{ccc}
\frac{h^3}{3\eta} & \frac{h^2\hat{a}}{3\eta} & 0 \\ 
\frac{h^2\hat{a}}{3\eta} & \frac{h\hat{a}^2}{3\eta}+D_a\hat{a} & 0 \\ 
0 & 0 & D_b b
\end{array} \right)\label{eq:Q_GD_mixture_substrate_reaction}
\end{equation}
with the dynamic viscosity $\eta$ and the diffusive coefficients \reviewchange{$D_{a}$ and $D_{b}$}. Note that \reviewchange{the} upper left $2\times 2$-block \reviewchange{in} (\ref{eq:Q_GD_mixture_substrate_reaction}) directly corresponds to the transport coefficients of the gradient dynamics formulation for thin films of mixtures and suspensions in the case without slip at the substrate \cite{ThTL2013prl}. In consequence, only diffusive transport is possible for species~$B$ (with mobility $D_b b$). The reactive flux of the adsorption-desorption process is given by the detailed balanced kinetics
\begin{equation}
J_r = r(h) \left(\exp\left[\frac{1}{k_b T}\frac{\delta F}{\delta b}\right]-\exp\left[\frac{1}{k_b T}\frac{\delta F}{\delta \hat{a}}\right]\right). \label{eq:adsorption_desorption_reactive_flux}
\end{equation}
We assume that the \reviewchange{rate function} $r(h)$ in (\ref{eq:adsorption_desorption_reactive_flux}) is film height-dependent to impose that adsorption and desorption may only occur where the macroscopic drop is in contact with the substrate. The rate takes the form
\begin{equation}
r(h) = r_0 \xi(h) = \frac{r_0}{2} \left[1+\tanh\left(\frac{h-h_0}{\delta h}\right)\right],\label{eq:r_film_height_dependent}
\end{equation}
where $r_0$ is the actual \reviewchange{(constant) rate} of the reaction and $\xi(h)$ is a smooth step-like function. The quantity $h_0$ is chosen slightly larger than the thickness of the equilibrium liquid adsorption layer (precursor film) and \reviewchange{$\delta h > 0$} is a measure for the steepness of the step that is chosen much smaller than the maximal drop height. Alternatively, the wetting energy may be adapted in such a way that no solute enters the adsorption layer. We stress that Eqs.~(\ref{eq:r_film_height_dependent}) conserves the gradient dynamics form, as (\ref{eq:adsorption_desorption_reactive_flux}) still expresses detailed balanced kinetics.
\subsubsection{Reduced two-field model with chemostatting}\label{sec:mixture_reduced_two_field_model}
In principle, the free energy (\ref{eq:F_drop_mixture_substrate_interaction}) can be supplemented by specific choices for $f, g_a$ and $g_b$ such that equations (\ref{eq:GD_mixture_h})-(\ref{eq:GD_mixture_b}) describe the time evolution of a particular system. However, here we further simplify (\ref{eq:GD_mixture_h})-(\ref{eq:GD_mixture_b}) by assuming that species~$A$ within the drop is kept at constant and uniform density $a_0$ by means of chemostatting. This may express that it is present in vast excess of what is needed to cover the substrate. Then, Eqs.~(\ref{eq:GD_mixture_h})-(\ref{eq:GD_mixture_b}) reduce to the two-field description
\begin{align}
\partial_t h &= \nabla\cdot\left[\frac{h^3}{3\eta}\nabla p\right],\label{eq:GD_mixture_reduced_h}\\
\partial_t b &= \nabla\cdot\left[D_b b \nabla \mu_b\right]-J_r,\label{eq:GD_mixture_reduced_b}
\end{align}
with the pressure 
\begin{align}
  p &= \partial_h f - \gamma\Delta h, \label{eq:mixture_reduced_p}
\end{align}
      and the chemical potential of $B$ 
\begin{align}
 \mu_b  &= g_b'+\partial_b f -\sigma_b \Delta b.\label{eq:mixture_reduced_mu_b}
\end{align}
The reactive flux becomes
\begin{equation}
J_r = r(h) \left(\exp\left[\frac{\mu_b}{k_b T}\right]-\exp\left[\frac{\mu_{a,0}}{k_b T}\right]\right),
\end{equation}
with $\mu_{a,0}$ as the chemostatted (uniform and constant) chemical potential of species~$A$ where $r(h)$ is still given by (\ref{eq:r_film_height_dependent}). In general, chemostatting breaks the gradient dynamics form and results in active rather than thermo\-dy\-na\-mic behavior. However, in the special case $\mu_{a,0}=0$ there is no energetic influx associated with chemostatting and dissipation remains unperturbed. In consequence, the system relaxes to thermodynamic equilibrium (defined by the \textit{separate} vanishing of all fluxes) and it can be shown that Eqs.~(\ref{eq:GD_mixture_reduced_h})-(\ref{eq:GD_mixture_reduced_b}) then represent a gradient dynamics (see \reviewchange{Appendix~\ref{app:GD_proof_reduced_model_mu_a_0}}).
\subsubsection{Free energy choices}
We now supplement the simplified model (\ref{eq:GD_mixture_reduced_h})-(\ref{eq:GD_mixture_reduced_b}) with specific choices for $f$ and $g_b$. The wetting energy is
\begin{equation}
f(h, b) = A_H\left(1+\lambda\frac{b}{b_0}\right)\left(-\frac{1}{2h^2}+\frac{h_a^3}{5h^5}\right), \label{eq:wetting_energy_expression}
\end{equation}
where $A_H$ is the Hamaker constant, $\lambda>0$ is a dimensionless proportionality factor, $b_0$ is some characteristic density of $B$, and $h_a$ denotes the thickness of the equilibrium liquid adsorption layer (precursor film). The choice (\ref{eq:wetting_energy_expression}) models a substrate that is rendered less wettable by the adsorbate and thus corresponds to the regime of partial wetting since the spreading coefficient $\reviewchange{S^*=f(h_a, b)}$ \cite{Genn1985rmp,BEIM2009rmp,Thie2010jpcm} is negative for all $b\geq 0$. Further, we employ the purely entropic local free energy
\begin{equation}
  g_b(b) = k_b T b \left[\ln\left(b/b_0\right) -1\right]
  \label{eq:entropic_free_energy_b}
\end{equation}
for species $B$. To study the model numerically, Eqs.~(\ref{eq:GD_mixture_reduced_h})-(\ref{eq:GD_mixture_reduced_b}) are nondimensionalized, see  \reviewchange{Appendix~\ref{app:chemically_sustained_wettability_gradients_nondimensional}}. From here on, all symbols denote dimensionless quantities. 
\subsubsection{Numerical results}\label{sec:numerical_results_running_drops}
\begin{figure}[tbh]
	\centering
	\includegraphics[width=0.485\textwidth]{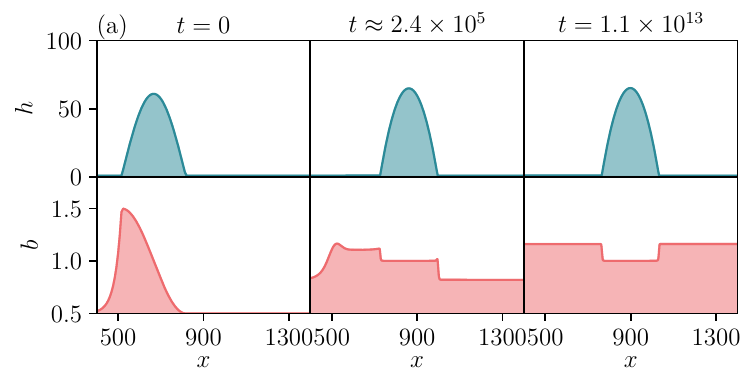}
	\includegraphics[width=0.485\textwidth]{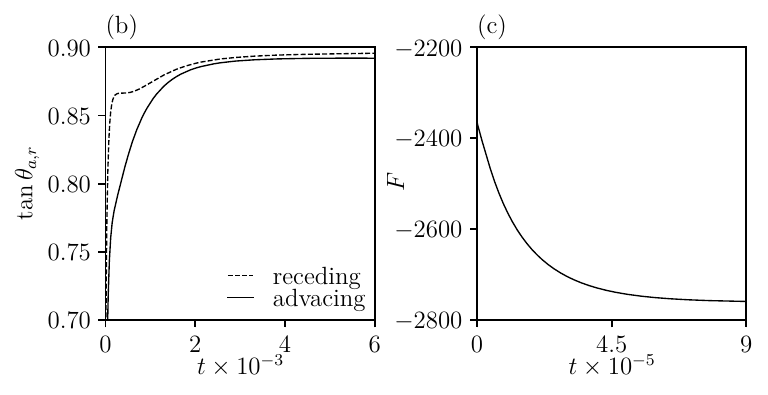}
\caption{(a) Snapshots of the film height profile $h$ and the adsorbate density profile $b$ at different times in the relaxational case at $\mu_{a,0}=0$. The drop first moves away from its initial position because initially at $t=0$ an asymmetric wettability profile is imposed via $b(x,0)$. Then, the drop slows down due to the equilibration of $b$. As equilibrium is approached, $b$ becomes \reviewchange{respectively} uniform inside and outside of the drop with interfaces in $b$ in the contact line regions.
(b) Dependence of the tangents of the advancing (solid line) and receding (dashed line) contact angles on time. Starting with an initially symmetric drop, the initial wettability gradient causes a difference in contact angles, with $\theta_a < \theta_r$. At the final equilibrium, $\theta_a = \theta_r$. (c)~The free energy $F$ decreases monotonically with time since the system is passive. The remaining parameters are $W=1, \lambda=0.5, r_0=0.0025, D_b=0.001, h_0 = 2$ and $\delta h = 0.3$. Only part of the computational domain $\Omega = \left[0, 2000\right]$ is shown}
\label{fig:run_undriven_plots.pdf}
\end{figure}

\begin{figure}[tbh]
	\centering
     \includegraphics[width=0.5\textwidth]{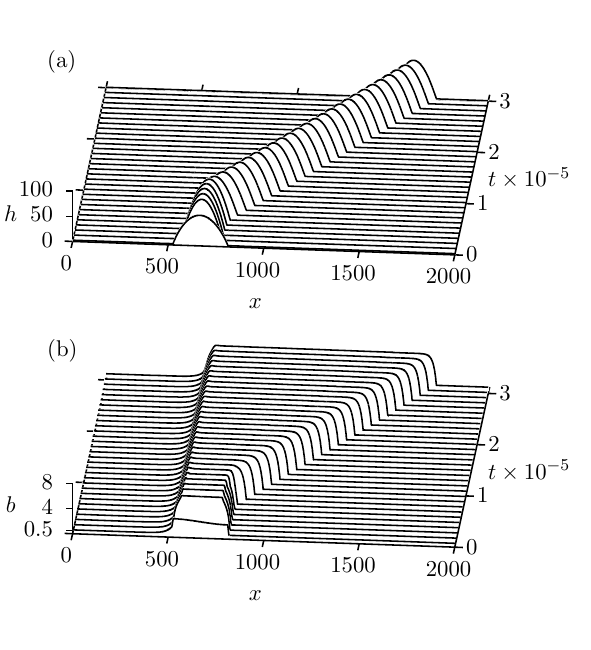}
     \includegraphics[width=0.35\textwidth]{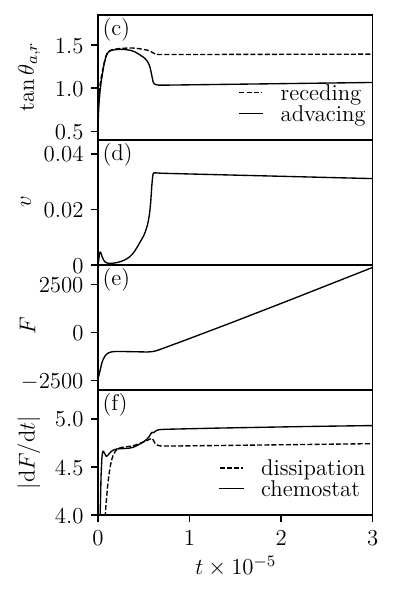}
\caption{Space-time plots illustrating the time evolution of (a) the film height profile $h$ and (b) the adsorbate density profile $b$ in the active case at $\mu_{a,0}=\ln 6$. The drop first moves due to an initial asymmetric wettability profile. After a short transient, the drop moves across the substrate at constant speed while self-sustaining the driving wettability gradient. (c) Dependence of the tangents of the advancing (solid line) and receding (dashed line) contact angles on time. After the acceleration phase, both contact angles remain distinct while the difference very slowly decreases. (d) Dependence of the drop's center-of-mass velocity $v$ on time. After the \reviewchange{initial} transient, $v$ slowly decreases due to adsorbate diffusion. (e) The free energy $F$ first shows a transient. Then, when the drop has reached its constant speed, $F$ increases linearly. (f) Shown are the corresponding energy rates associated with dissipation $R_\text{diss}$ (dashed line) and chemostatting $R_\text{chem}$ (solid line). Note that $R_\text{diss}\ge0$ per definition. After the transient, the chemostat supplies energy at constant rate while the system constantly dissipates energy. The remaining parameters are as in Fig.~\ref{fig:run_undriven_plots.pdf}. The computational domain is $\Omega = \left[0, 2000\right]$}
\label{fig:run_driven_plots}
\end{figure}

To elucidate some general features of the reduced model (\ref{eq:GD_mixture_reduced_h})-(\ref{eq:GD_mixture_reduced_b}), we perform time simulations of the nondimensional equations (\ref{eq:GD_mixture_reduced_h_nondimensional})-(\ref{eq:GD_mixture_reduced_b_nondimensional}) using the finite-element library \textsc{oomph-lib} \cite{HeHa2006} with periodic boundary conditions on the one-dimensional domain $\Omega = \left[0, 2000\right]$. Further details on the numerical methods are given in Appendix~\ref{app:numerical_details}. We consider the passive \reviewchange{(relaxational)} case at $\mu_{a,0}=0$ and the active (chemostat-driven) case at $\mu_{a,0}\neq 0$, while all other parameters are identical. In both cases, we initialize the simulation with a symmetric drop profile and an asymmetric concentration profile, i.e., we impose an initial wettability gradient. In the passive case (see Fig.~\ref{fig:run_undriven_plots.pdf}), we find that the drop initially moves away from its starting position driven by the initial wettability gradient. However, it \reviewchange{subsequently slows down} due to the equilibration of $b$ (Fig.~\ref{fig:run_undriven_plots.pdf}a). Note that also the wettability to the right of the drop slowly decreases because $B$ slowly diffuses outside the drop (\reviewchange{and} across the periodic domain boundaries). In the final equilibrium state, the concentration of~$B$ is \reviewchange{respectively} uniform inside and outside the drop with a steep gradient located in the contact line region. The difference in coverage can be traced back to the reduction of the chemical potential of $B$ outside the drop due to the concentration-dependent wetting energy. We note that the equilibrium value of $b$ inside the drop is determined as `bare' reactive equilibrium, i.e., it must be unity in dimensionless units (dimensional $b=b_0$). The equilibration process is also visible in Fig.~\ref{fig:run_undriven_plots.pdf}b that shows the dependence of the advancing and receding contact angles on time. They approach the same equilibrium value. As the system is passive, the free energy~$F$ decreases monotonically (Fig.~\ref{fig:run_undriven_plots.pdf}c).\footnote{While we only show part of the time evolution of the free energy, we have checked that $F$ always decreases until the final equilibrium state is reached.}

Fig.~\ref{fig:run_driven_plots} shows an example for the active case with $\mu_{a, 0}>0$. There, we find continuously moving drops: After a short transient, the drop moves along the substrate at nearly constant speed maintained by the self-sustained wettability gradient across its length. Note that the drop's center-of-mass velocity $v$ slightly decreases with time (Fig.~\ref{fig:run_driven_plots}d) due to slow diffusion of $b$ across the periodic domain boundaries which reduces the wettability in front of the drop.\footnote{A similar effect occurs for no-flux boundary conditions, where some $B$ diffuses in front of the drop. For boundary conditions which allow for constant $v$, see \cite{ThJB2004prl}.} The advancing and receding dynamic contact angles ($\theta_{a}$ and $\theta_{r}$) differ when the drop moves (Fig.~\ref{fig:run_driven_plots}c) with the difference slightly decreasing with time due to the mentioned diffusion. In contrast to the passive case, after the transient the free energy increases approximately linearly with time (Fig.~\ref{fig:run_driven_plots}e). Considering the rate of energy influx due to the chemostat $R_\text{chem}$ and the dissipation rate $R_\text{diss}$ \reviewchange{(an energy outflux)} during the sustained drop motion (Fig.~\ref{fig:run_driven_plots}f, also see \reviewchange{Appendix~\ref{app:energetic_influx_dissipation_active}}), both rates are roughly constant since there is a continuous turnover of suspended particles. Namely, the chemostat replaces particles in the drop that adsorb onto the substrate. Since the wettability profile across the drop remains roughly stationary in the comoving frame, the chemostat must supply energy at a constant rate. Similarly, energy is dissipated at a constant rate. Therefore, because chemostatting outweighs dissipation ($R_\text{chem} > R_\text{diss}$), the free energy increases linearly, reflecting the continuous coating of the substrate. Strikingly, the relative difference $(R_\text{chem}-R_\text{diss})/R_\text{chem}$ between the rates of chemostatting and dissipation is less than five percent, i.e., only a small part of the energy from the chemostat remains in the system while a much larger part is immediately dissipated. 

\subsection{Reactive surfactants -- Sessile drops covered by autocatalytic surfactants}\label{sec:drops_with_autocatalysis} 
\begin{figure}[tbh]
	\centering
	\includegraphics[width=0.8\textwidth]{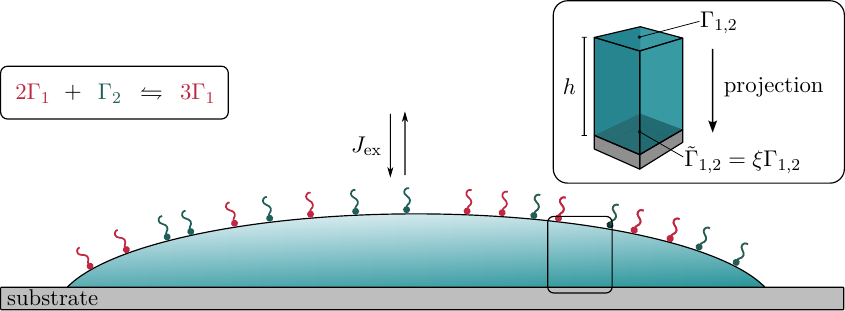}
\caption{Sketch of the considered geometry.  A shallow drop of a partially wetting liquid on a flat solid substrate is covered by two species of insoluble surfactant which engage in an autocatalytic conversion reaction. Additionally, the two species are in contact with an external bath that acts as a chemostat and exchanges surfactant with the drop surface ($J_\text{ex}$). The film height profile is described by $h$ and the surfactant density profiles on the film surface by \reviewchange{$\Gamma_{1}$ and $\Gamma_{2}$}. For a gradient dynamics description, the densities have to be projected onto the substrate plane where the projected densities are \reviewchange{$\tilde{\Gamma}_{\alpha}=\xi\Gamma_{\alpha}$ with $\alpha=1,2$}. Here, $\xi$ is the metric factor of the film surface}
\label{fig:sketch_drop_reactive_surfactants}
\end{figure}

Our second example are shallow drops of partially wetting liquid that are covered by two species of insoluble reactive surfactants (Fig.~\ref{fig:sketch_drop_reactive_surfactants}). In general, systems involving reactive surfactants are usually treated starting from a \reviewchange{hydrodynamic} perspective where chemical reactions are added in an \textit{ad hoc} manner \cite{DaPi1984jcis,BuRV1993jcis,ReRV1994pf,DMMB1996prl,PTTK2007pf,ChSD2022pre}. Here, we employ the described  gradient dynamics approach to \reviewchange{model} the interplay between autocatalytic chemical reactions, Marangoni effects and wetting physics in a thermodynamically consistent manner, i.e., with an appropriate passive limit.
In particular, we assume that the two species engage in the autocatalytic reaction
\begin{equation}
2\Gamma_1+\Gamma_2 \leftrightharpoons 3\Gamma_1,\label{eq:surfactant_autocatalysis}
\end{equation}
where $\Gamma_{1}$ and $\Gamma_{2}$ denote the particle densities of the two species (particles per unit interface area). Additionally, we allow for surfactant exchange with an external chemostatted `bath'.

The free energy of the system is given by
\begin{equation}
F = \int_\Omega\left[f(h)+\xi g(\Gamma_1, \Gamma_2)\right]\text{d}^2 r,
\end{equation}
where $f(h)$ is the wetting energy and $g(\Gamma_1, \Gamma_2)$ is the energy density of the surfactant-covered free surface. For simplicity, here, we neglect a possible surfactant dependence of the wetting energy and, similarly, a dependence of the surface energy on the film height \cite{TSTJ2018l}. Note that we use the exact metric factor $\xi = \sqrt{1+\vert\nabla h\vert^2}$ and not its long-wave approximation that was used in Section~\ref{sec:drops_with_wettability_gradient}. To compute the correct pressure and chemical potentials from $F$ we need to identify the independent fields that correspond to the local liquid volume and the local particle \reviewchange{numbers} of surfactant in a small liquid column. In analogy to the argument in Section~\ref{sec:drops_with_wettability_gradient} for the bulk concentration, the independent fields are the local film height $h$ and the projected densities $\tilde{\Gamma}_\alpha = \xi\Gamma_\alpha$ (particles per unit substrate area) \reviewchange{with $\alpha=1,2$}. For further discussion, we refer to \cite{ThAP2012pf,ThAP2016prf}. The resulting pressure and chemical potentials are 
\begin{align}
p &= \frac{\delta F}{\delta h} = \partial_hf - \nabla\cdot\left[\frac{1}{\xi}\left(g-\Gamma_1\partial_{\Gamma_1}g-\Gamma_2\partial_{\Gamma_2}g\right)\nabla h\right],\label{eq:autocatalytic_surfactants_pressure} \\
\mu_\alpha &= \frac{\delta F}{\delta\tilde{\Gamma}_\alpha} = \partial_{\Gamma_\alpha} g, \label{eq:autocatalytic_surfactants_chemical_potentials}
\end{align}
respectively. The coupled time evolution equations are
\begin{align}
	\partial_t h &= \nabla\cdot\left[Q_{hh}\nabla\frac{\delta F}{\delta h}+Q_{h\Gamma_1}\nabla\frac{\delta F}{\delta \tilde{\Gamma}_1}+Q_{h\Gamma_2}\nabla\frac{\delta F}{\delta \tilde{\Gamma}_2}\right],\label{eq:GD_autocatalytic_surfacants_h}\\
	\partial_t \tilde{\Gamma}_1 &= \nabla\cdot\left[Q_{\Gamma_1h}\nabla\frac{\delta F}{\delta h}+Q_{\Gamma_1\Gamma_1}\nabla\frac{\delta F}{\delta \tilde{\Gamma}_1}+Q_{\Gamma_1\Gamma_2}\nabla\frac{\delta F}{\delta \tilde{\Gamma}_2}\right]+J_r+J_{\text{ex},1},\label{eq:GD_autocatalytic_surfacants_g1}\\
	\partial_t \tilde{\Gamma}_2 &= \nabla\cdot\left[Q_{\Gamma_2h}\nabla\frac{\delta F}{\delta h}+Q_{\Gamma_2\Gamma_1}\nabla\frac{\delta F}{\delta \tilde{\Gamma}_1}+Q_{\Gamma_2\Gamma_2}\nabla\frac{\delta F}{\delta \tilde{\Gamma}_2}\right]-J_r+J_{\text{ex},2},\label{eq:GD_autocatalytic_surfacants_g2}
\end{align}
where the symmetric positive definite mobility matrix is given by
\begin{equation}
	\tens{Q} = \left(\begin{array}{ccc}
Q_{hh} & Q_{h\Gamma_1} & Q_{h\Gamma_2} \\ 
Q_{\Gamma_1h} & Q_{\Gamma_1\Gamma_1} & Q_{\Gamma_1\Gamma_2} \\ 
Q_{\Gamma_2h} & Q_{\Gamma_2\Gamma_1} & Q_{\Gamma_2\Gamma_2}
\end{array} \right) = \left(\begin{array}{ccc}
	\frac{h^3}{3\eta} & \frac{h^2\Gamma_1}{2\eta} & \frac{h^2\Gamma_2}{2\eta} \\ 
	\frac{h^2\Gamma_1}{2\eta} & \frac{h\Gamma_1^2}{\eta}+D_1\Gamma_1 & \frac{h\Gamma_1\Gamma_2}{\eta}\\ 
	\frac{h^2\Gamma_2}{2\eta} &  \frac{h\Gamma_1\Gamma_2}{\eta}  &  \frac{h\Gamma_2^2}{\eta}+D_2\Gamma_2
	\end{array} \right). \label{eq:autocatalyic_surfactants_Q}
\end{equation}
It represents a straightforward generalization of the mobilities for the single-surfactant case \cite{ThAP2012pf} and contains advective and diffusive contributions. Here, $D_\alpha > 0$ are constant diffusive mobilities. For the reactive flux, we have
\begin{equation}
J_r = r\left[\exp\left(\frac{2}{k_b T}\frac{\delta F}{\delta \tilde{\Gamma}_1}+\frac{1}{k_b T}\frac{\delta F}{\delta \tilde{\Gamma}_2}\right)-\exp\left(\frac{3}{k_b T}\frac{\delta F}{\delta \tilde{\Gamma}_1}\right)\right], \label{eq:GD_autocatalysis_terms}
\end{equation}
where $r>0$ is the \reviewchange{constant rate}. For simplicity, the surfactant exchange with the external chemostatted bath is assumed to be proportional to the chemical potential difference between the surfactants on the film and in the bath, i.e.,
\begin{align}
J_{\text{ex},1} = -\beta_1\left(\frac{\delta F}{\delta \tilde{\Gamma}_1}-\mu_{0,1}\right) = -\beta_1\frac{\delta F}{\delta \tilde{\Gamma}_1}-\beta\mu, \label{eq:exchange_flux_1}\\
J_{\text{ex},2} = -\beta_2\left(\frac{\delta F}{\delta \tilde{\Gamma}_2}-\mu_{0,2}\right) = -\beta_2\frac{\delta F}{\delta \tilde{\Gamma}_2}+\beta\mu. \label{eq:exchange_flux_2}
\end{align}
Here, $\beta_\alpha>0$ are transition rates for the transfer between the ambient bath and the film surface and the $\mu_{0,\alpha}$ are the imposed chemical potentials of the two species in the bath. Here, we restrict our attention to part of the possible parameter space by assuming the two terms $\beta_\alpha\mu_{0\alpha}$ to be equal in magnitude (given by $\beta\mu$) and opposite in sign. With $\beta\mu > 0$ and the sign choices as in (\ref{eq:exchange_flux_1})-(\ref{eq:exchange_flux_2}), $\Gamma_1$  tends to transfer from the film surface to the bath, with an inverse tendency for $\Gamma_2$. The parameter $\beta\mu$ then represents the sustained nonequilibrium driving force of the chemostat. For $\beta\mu=0$ there is no energy flux in or out of the system associated with chemostatting. In this case, the system is passive and equations (\ref{eq:GD_autocatalytic_surfacants_h})-(\ref{eq:GD_autocatalytic_surfacants_g2}) again represents a gradient dynamics.

Since the \reviewchange{dynamic equations} are expressed in terms of the projected densities, the nonconserved terms explicitly depend on the metric factor of the free surface. This corresponds to a basic form of geometry-induced chemo-mechanical coupling which, here, leads to an overall slowing-down of the reaction flux (\ref{eq:GD_autocatalysis_terms}) on the  convex drop surface. While the metric factor only weakly deviates from unity in the long-wave limit and can then often be neglected, on strongly curved surfaces and in bulk fluids chemo-mechanical couplings can give rise to pattern forming instabilities and sustained oscillations/motion \cite{MJK2019prl, WuGF2023pre, BoJG2011prl, GrKG2017arb}.
\subsubsection{Free energy choices and further simplifications}
Starting from the gradient dynamics form (\ref{eq:GD_autocatalytic_surfacants_h})-(\ref{eq:GD_autocatalytic_surfacants_g2}), we now supplement the model with specific energies and introduce further simplifications. We assume that the surfactant species only sparsely cover the film surface, implying purely entropic contributions to the surface free energy, i.e.,
\begin{equation}
g_s\left(\Gamma_1, \Gamma_2\right) = \gamma^0 + k_b T\,\Gamma_1\left[\ln\left(\Gamma_1 a_1^2\right)-1\right]+ k_b T\,\Gamma_2\left[\ln\left(\Gamma_2 a_2^2\right)-1\right].
\end{equation}
Here, $\gamma^0$ is the surface tension of the bare surface and $a_{1}, a_{2}>0$ are characteristic length scales of the two surfactants. Additionally, we use the simple wetting energy
\begin{equation}
f(h) = A_H\left(-\frac{1}{2h^2}+\frac{h_a^3}{5h^5}\right), \label{eq:wetting_energy_expression_simple}
\end{equation}
where $A_H$ is the Hamaker constant and $h_a$ is the equilibrium adsorption layer height (precursor thickness), corresponding to the minimum of $f(h)$. Again, we consider the case of partial wetting, i.e., of negative spreading coefficient $\reviewchange{S^* = f(h_a)<0}$ \cite{Genn1985rmp,BEIM2009rmp,Thie2010jpcm}. The resulting pressure and chemical potentials are
\begin{align}
p &= \partial_h f- \nabla\cdot\left[\tfrac{1}{\xi}\left(\gamma^0 -k_b T \,\Gamma_1 -k_b T\, \Gamma_2\right)\nabla h\right],\label{eq:p_autocalytic_surfactants} \\
\mu_\alpha &= k_b T \ln(\Gamma_\alpha a_\alpha^2). \label{eq:mu_autocalytic_surfactants}
\end{align}
\reviewchange{Finally, we assume that $\xi\approx 1$,~i.e., we transition from the full-curvature formulation to the long-wave limit ($\vert\vert \nabla h \vert\vert \ll 1$). We further assume} that the changes in surface tension due to the presence of surfactant are small compared to $\gamma^0$. Then the pressure reduces to
\begin{equation}
p = \partial_h f - \gamma^0\Delta h. \label{eq:p_simplified_autocatalytic_surfactants}
\end{equation}
This approximation does not affect the Marangoni flows but only simplifies the Laplace pressure. It is commonly used when considering Marangoni effects \cite{OrDB1997rmp,CrMa2009rmp}.
The resulting time evolution equations are
\begin{align}
\partial_t h =&\: \nabla\cdot \left[\frac{h^3}{3\eta}\nabla p\right]+k_b T\nabla\cdot\left[\frac{h^2}{2\eta}\nabla\Gamma_1\right]+k_b T\nabla\cdot\left[\frac{h^2}{2\eta}\nabla\Gamma_2\right],
\label{eq:h_hydrodynamic_autocatalysis}\\[2ex]
\begin{split}
\partial_t \Gamma_1 =&\: \nabla\cdot \left[\frac{h^2\Gamma_1}{2\eta}\nabla p\right]+k_b T\nabla\cdot\left[\left(\frac{h\Gamma_1}{\eta}+D_1\right)\nabla\Gamma_1\right]+k_b T\nabla\cdot\left[\frac{h\Gamma_1}{\eta}\nabla\Gamma_2\right]\\
&+r\left[\left(\Gamma_1 a_1^2\right)^2\left(\Gamma_2 a_2^2\right)-\left(\Gamma_1 a_1^2\right)^3\right]-\beta_1 k_b T\ln(\Gamma_1 a_1^2)-\beta\mu,
\end{split} \label{eq:g1_hydrodynamic_autocatalysis}
\\[2ex]
\begin{split}
\partial_t \Gamma_2 =&\: \nabla\cdot \left[\frac{h^2\Gamma_2}{2\eta}\nabla p\right]+k_b T\nabla\cdot\left[\frac{h\Gamma_2}{\eta}\nabla\Gamma_1\right]+k_b T\nabla\cdot\left[\left(\frac{h\Gamma_2}{\eta}+D_2\right)\nabla\Gamma_2\right]\\
&-r\left[\left(\Gamma_1 a_1^2\right)^2\left(\Gamma_2 a_2^2\right)-\left(\Gamma_1 a_1^2\right)^3\right]-\beta_2 k_b T\ln(\Gamma_2 a_2^2)+\beta\mu.
\end{split}\label{eq:g2_hydrodynamic_autocatalysis}
\end{align}
To numerically study the model, we nondimensionalize (\ref{eq:h_hydrodynamic_autocatalysis})-(\ref{eq:g2_hydrodynamic_autocatalysis}). The rescaled model as well as details on the choice of scales and dimensionless parameters can be found in \reviewchange{Appendix~\ref{app:insoluble_autocatalytic_surfactants_nondimensional}}. In the following, all symbols denote dimensionless quantities.
\subsubsection{Numerical results}\label{sec:numerical_results_surfactants}

\begin{figure}[tbh]
	\centering
	\includegraphics[width=0.85\textwidth]{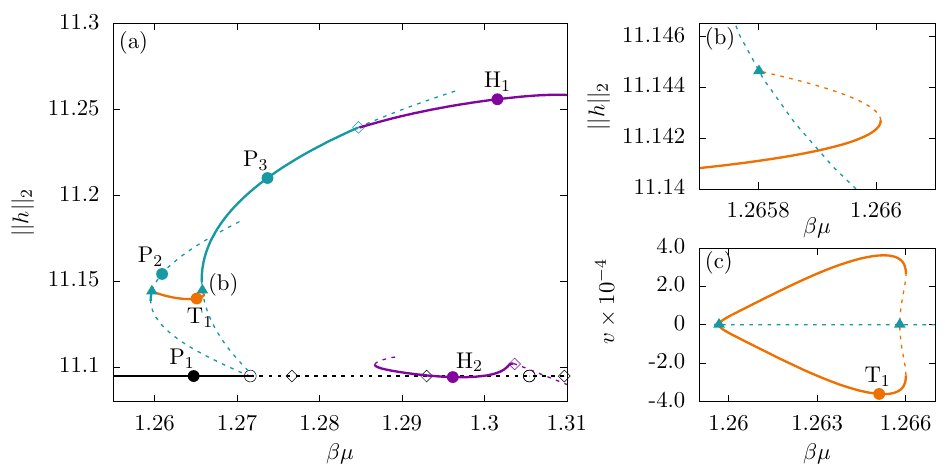} 
        \caption{(a) Partial bifurcation diagram for drop states in dependence of the chemostat driving strength $\beta\mu$. The solution measure corresponds to the norm $||h||_2$. Shown are \reviewchange{branches of linearly stable states (solid lines) and directly related sections of the branches of unstable states (dashed lines).} The horizontal black line represents the base state - the simple resting drop with homogeneous surfactant concentration. \reviewchange{Cyan} lines correspond to steady drops with localized surfactant Turing-like patterns, orange lines are branches of steadily traveling drops and purple ones represent standing wave-like drop oscillations. Labeled filled circles denote selected states shown in Fig.~\ref{fig:bifdiag_solution_profiles}. Panel~(b) magnifies the region where a supercritical drift-pitchfork \reviewchange{bifurcation} occurs resulting in the emergence of the first two branches of traveling drops (left- and right traveling, otherwise identical states). The corresponding drop velocities $v$ are given in panel (c) \reviewchange{for the entire branch}. The remaining parameters are $W=3, \delta=1, D_1=1.4, D_2=0.1, r=0.8, \beta_1=1$ and $\beta_2=0.5$}
\label{fig:drop_autocatalysis_bifdiag_betamu}
\end{figure}

\begin{figure}[tbh]
	\centering
	\includegraphics[width=0.8\textwidth]{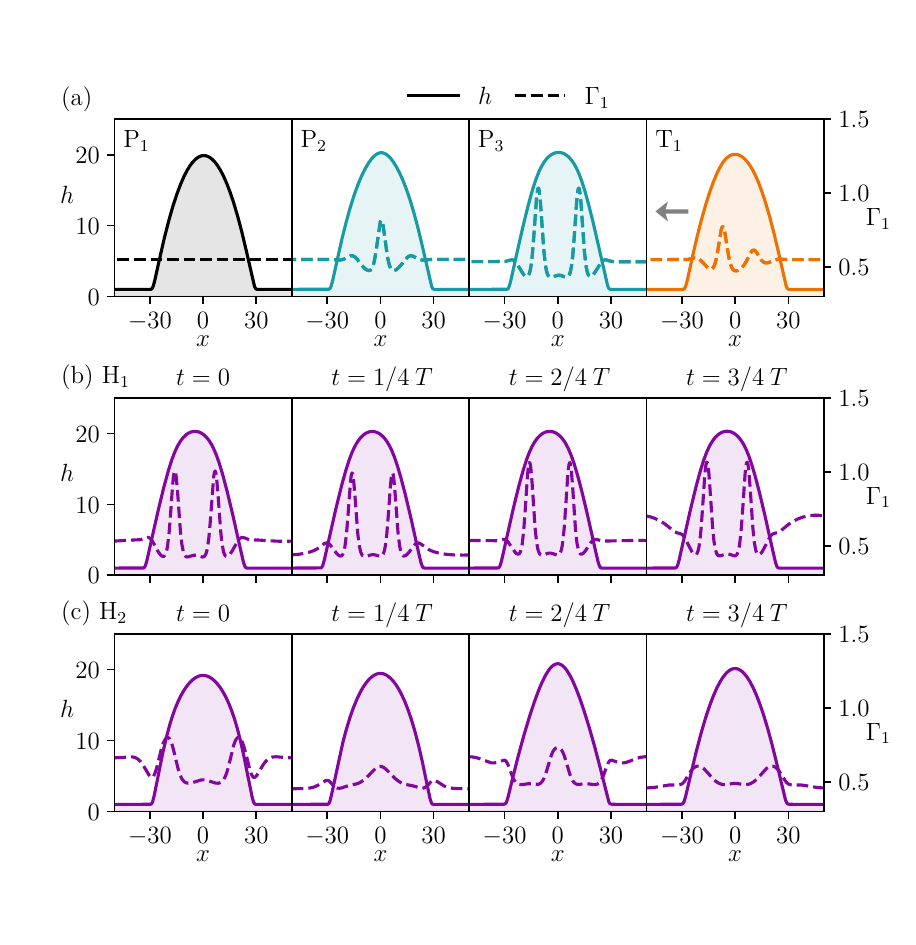}
\caption{(a) \reviewchange{Selected profiles of the film height $h$ \reviewchange{(solid lines)} and the surfactant $\Gamma_1$ \reviewchange{(dashed lines)} corresponding to steady (P$_1$ to P$_3$) and traveling (T$_1$) drop states at the respective points marked in Fig.~\ref{fig:drop_autocatalysis_bifdiag_betamu}a. The steadily traveling drop is shown in the comoving frame with speed $v$, the arrow indicates the direction of travel. (b),~(c) \reviewchange{Corresponding profiles of standing wave-like drop-surfactant oscillations at different times within one temporal period at the points H$_1$ and H$_2$ in Fig.~\ref{fig:drop_autocatalysis_bifdiag_betamu}a. The periods are (b) $T\approx 10.58$ and (c) $T\approx 20.71$}}}
\label{fig:bifdiag_solution_profiles}
\end{figure}

\begin{figure}[tbh]
	\centering
	\includegraphics[width=0.8\textwidth]{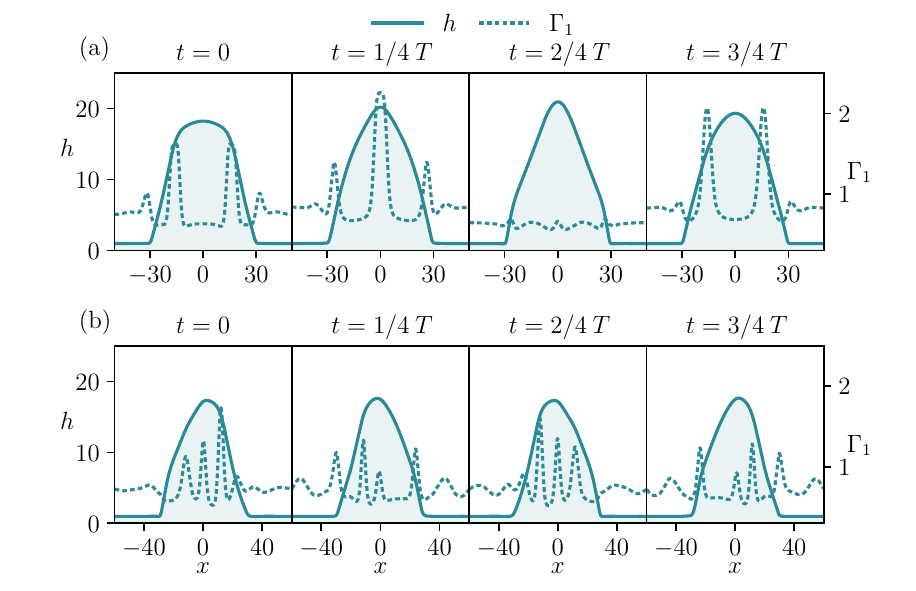}
        \caption{\reviewchange{Profiles of the film height $h$ (solid lines) and surfactant $\Gamma_1$ (dashed lines) of standing wave-like drop-surfactant oscillations corresponding to (a) a stable breathing mode and (b) an unstable swaying mode at different times within one temporal period. The parameters are (a) $W=4, \delta=1, D_1=2, D_2=0.15, r=1.1, \beta_1 =6, \beta_2=1$ and $\beta\mu=5.3992$ for a drop volume $V\approx 816$ and domain size $L=100$, and (b) $W = 1.504, \delta=1, D_1=4, D_2=0.3, r=1.2,\beta_1=6, \beta_2=1$ and $\beta\mu = 5.7868$ for $V=720$ and $L=120$. The periods are (a) $T\approx 6.04$ and (b) $T\approx 9.28$}}
\label{fig:oscillatory_modes}
\end{figure}

\begin{figure}[tbh]
	\centering
	\includegraphics[width=0.9\textwidth]{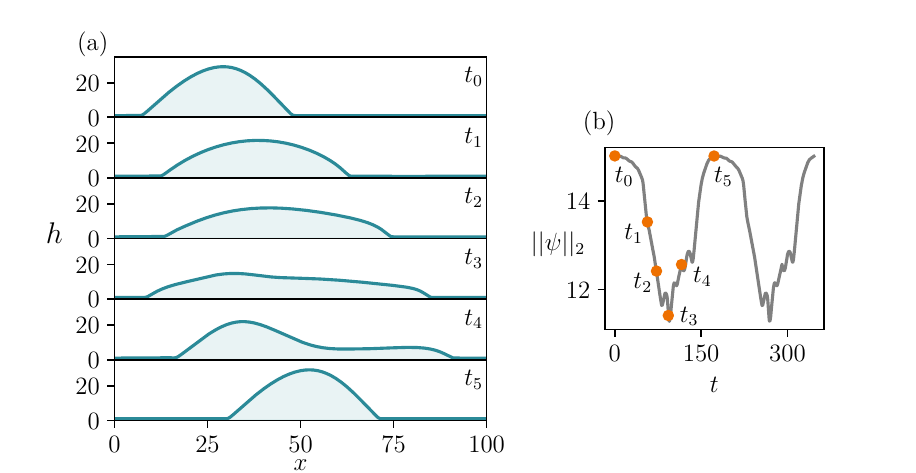}
\caption{\reviewchange{(a) Series of drop profiles at times $t_0$ to $t_5$ depicting drop expansion and contraction for a stable modulated traveling wave-like `crawling' drop-surfactant state in the laboratory frame at $W=7, D_1=1.3, D_2=0.1, \delta=1, r=0.8, \beta_1 = 1.4, \beta_2 = 0.7$, $\beta\mu=2.2$, $V\approx 816$, $L=100$. (b) The L$_2$-norm $\vert\vert\psi\vert\vert_2 = \sqrt{1/L\int_0^L\left(h^2+p^2+\Gamma_1^2+\Gamma_2^2\right)\:\text{d}x}$ as a function of time. Because the L$_2$-norm is invariant w.r.t.\ state translation, the observed modulated traveling wave results in time-periodic behavior. Markers represent the drop profiles from (a)}}
\label{fig:modulated_running_drop}
\end{figure}

\begin{table}[tbh]
\renewcommand\arraystretch{0.8}
\setlength{\tabcolsep}{5pt}
\centering
\caption{Summary of the \reviewchange{symbols used to mark different bifurcation types}}
\begin{tabular}{lcc}
\toprule[1pt]
Symbol & Base branch & Bifurcation type\\
\midrule
\hspace{0.625cm}\raisebox{0.125cm}{\circle{0.2cm}} & steady state & pitchfork \\
\midrule[0.05pt]
\multirow{2}{*}{\hspace{0.5cm}$\diamond$} & steady state & Hopf \\
 & standing wave & torus \\ 
\midrule[0.05pt]
\hspace{0.5cm}$\blacktriangle$ & steady state & drift-pitchfork\\
\bottomrule[1pt]
\end{tabular}
\label{tab:bifurcations_convention}
\end{table}

We now show a few typical features of chemo-mechanical drop dynamics described by Eqs.~(\ref{eq:h_hydrodynamic_autocatalysis})-(\ref{eq:g2_hydrodynamic_autocatalysis}). We focus on two-dimensional drops, i.e., one-dimensional substrates, and analyze the nondimensional model (\ref{eq:h_hydro_autocatal_nondim})-(\ref{eq:g2_hydro_autocatal_nondim}) given in \reviewchange{Appendix~\ref{app:insoluble_autocatalytic_surfactants_nondimensional}}. In our analysis we use numerical path continuation, employing the continuation package $\textsc{pde2path}$ \cite{UeWR2014nmma}, supplemented by time simulations obtained with the finite-elements library $\textsc{oomph-lib}$ \cite{HeHa2006}.  \reviewchange{The latter are used to obtain starting states needed to initiate the continuation of time-periodic states and to study states that are not easily accessible by continuation}. Further details are given in Appendix~\ref{app:continuation}. For the bifurcation diagram we use as solution measure the $L_2$-norm of the film height
\begin{equation}
\vert\vert h\vert\vert_2 = \sqrt{\frac{1}{L}\int_{-L/2}^{L/2} h^2\:\text{d}x}
\end{equation}
with the domain size $L$. For time-periodic states we use the time-averaged norm $1/T \int_0^T \vert\vert h\vert\vert_2\:\text{d}t$, where $T$ is the temporal period. We use the chemostat driving strength $\beta\mu$ as main control parameter and investigate how the drop behavior changes as the system is driven further away from thermodynamic equilibrium. Most results are obtained for a periodic domain of size $L=100$ and a drop volume of $V=\int_{-L/2}^{L/2}h\:\text{d}x\approx 816$, corresponding to small drops with a maximal height of about 20 times the adsorption layer height.  
In the bifurcation diagram we indicate different bifurcations using the specific symbols listed in Table~\ref{tab:bifurcations_convention}.

Fig.~\ref{fig:drop_autocatalysis_bifdiag_betamu}a shows the resulting bifurcation diagram in dependence of the driving strength $\beta\mu$. Only a selected part of the rich \reviewchange{structure} is shown, namely, branches of linearly stable resting, traveling and oscillating drop states and relevant sections of directly related \reviewchange{branches of unstable states}. For clarity, we do not show less relevant branches of unstable states.
Selected corresponding drop and concentration profiles are given in Fig.~\ref{fig:bifdiag_solution_profiles}.

The horizontal line in Fig.~\ref{fig:drop_autocatalysis_bifdiag_betamu}a corresponds to the base state -- a simple resting sessile drop with homogeneous surfactant concentration. It is the only shown state that already exists in the passive limit ($\beta\mu=0$). As $\beta\mu$ is increased, the simple drop becomes unstable in a double pitchfork bifurcation where two branches of unstable steady drop states with localized Turing-like surfactant patterns emerge subcritically (\reviewchange{cyan} lines).
The first patterned branch (once unstable) is then rendered linearly stable in a saddle-node bifurcation and shortly after is again destabilized at a drift-pitchfork bifurcation (filled triangle) \cite{GGGC1991pra,FaDT1991jpi}. There, a pair of branches of steadily traveling drops emerges (orange line) that connect the first  patterned branch to the second one (doubly unstable). The corresponding drop velocities are given in \reviewchange{Fig.~\ref{fig:drop_autocatalysis_bifdiag_betamu}c}. Note that the traveling drops are mostly stable states but become  unstable in a saddle-node bifurcation shortly before they connect to the second patterned branch in \reviewchange{another} drift-pitchfork bifurcation, see Fig.~\ref{fig:drop_autocatalysis_bifdiag_betamu}b. Beyond this bifurcation, the second patterned branch stabilizes in a saddle-node bifurcation, remains linearly stable for an extended $\beta\mu$-range, before it becomes unstable at a Hopf bifurcation, and is here not further discussed.
There, a branch of time-periodic states emerges that correspond to standing wave-like oscillations.\footnote{Due to the relaxed numerical tolerance and mesh resolution necessary for the continuation of this branch, \reviewchange{its end does not exactly match the location of the Hopf point of the branch of steady patterns.} However, we have checked that the profiles and periods match at the  bifurcation and have therefore shifted the L$_2$-norm of the branch by $0.00165$ for presentational purposes.} The stable oscillatory state (H$_1$) is illustrated in Fig.~\ref{fig:bifdiag_solution_profiles}b. The oscillation in the surfactant pattern is relatively weak and the one of the drop shape is not visible on the scale of the figure, indicating a weak coupling between chemical reactions and bulk liquid. Remarkably, there exists a range of driving strengths where the simple drop branch, the second patterned branch and the branch of steadily traveling drops are simultaneously linearly stable.

Time simulations at larger values of $\beta\mu$ also result in time-periodic states with noticeable surfactant-bulk coupling (cf.~state H$_2$ in Fig.~\ref{fig:drop_autocatalysis_bifdiag_betamu}a, profiles in Fig.~\ref{fig:bifdiag_solution_profiles}c) leading to left-right symmetric drop oscillations, i.e., a breathing mode. Following the corresponding branch by continuation, we find that it destabilizes in a saddle-node bifurcation at a smaller value of $\beta\mu$ and in a torus bifurcation at a larger $\beta\mu$. The emerging linearly stable quasi-periodic drop-surfactant oscillations are also found in time simulations initialized just beyond this bifurcation (not shown).

Despite its already rather complex structure, the bifurcation diagram in Fig.~\ref{fig:drop_autocatalysis_bifdiag_betamu}a contains general features that are also found at stronger surfactant-bulk coupling, however, within much more involved bifurcation structures (not shown). Seemingly universal key elements are the emergence of patterned surfactant branches in double pitchfork bifurcations from the simple drop branch, the pairwise interconnection of these patterned branches by branches of steadily traveling drops as `rung states' and the virtually exclusive existence of quasi-periodic states at large driving strengths. Furthermore, in general, we find two modes of drop oscillation, breathing and swaying drops, which are illustrated in Fig.~\ref{fig:oscillatory_modes}. For breathing modes (Fig.~\ref{fig:oscillatory_modes}a), the film height profile and the surfactant distributions are left-right symmetric at all times. Swaying modes (Fig.~\ref{fig:oscillatory_modes}b) are generally antisymmetric, i.e., film height and surfactant profiles that are offset by half a temporal period are related by reflection. Both modes normally emerge in Hopf bifurcations. Lastly, in time simulations we also find stable modulated traveling drops that represent a more complicated form of transport (Fig.~\ref{fig:modulated_running_drop}). Such states may for instance emerge from drift-pitchfork bifurcations of standing wave-like states or from Hopf bifurcations of steadily traveling drop states. In time simulations, the drop begins to drift after an initial spontaneous breaking of the left-right symmetry and it oscillates asymmetrically while moving. In the case of Fig.~\ref{fig:modulated_running_drop}a, the drop periodically \reviewchange{flattens and} forms a large protrusion in the direction of propagation and subsequently contracts again to a cap-like shape, resulting in net movement of the whole drop. The periodicity in the comoving frame is best appreciated in the dependence of the L$_2$-norm $\vert\vert\psi\vert\vert_2 = \sqrt{1/L\int_0^L\left(h^2+p^2+\Gamma_1^2+\Gamma_2^2\right)\:\text{d}x}$ on time (Fig.~\ref{fig:modulated_running_drop}b). This measure is minimal and maximal when the drop's protrusion is at maximal length \reviewchange{(fully flattened, see $t_3$ in Fig.~\ref{fig:modulated_running_drop}a)} and when the drop is nearly cap-like \reviewchange{(fully contracted, see $t_0, t_5$ in Fig.~\ref{fig:modulated_running_drop}a)}, respectively. 

\section{Conclusion}\label{sec:conclusion}
We have introduced a way to write reactive mesoscopic hydrodynamics (reactive thin-film models) in gradient dynamics form. On the one hand, this thermodynamically consistent description \reviewchange{models} passive processes where the system ultimately approaches thermodynamic equilibrium. On the other hand, it allows for a controlled introduction of persistent nonequilibrium driving by making the dynamical couplings nonreciprocal (breaking the Onsager relations of transport coefficients and introducing indefinite transport matrices \cite{FHKG2023pre, GLFT2024arxiv}, breaking detailed balance of chemical reactions \cite{DGMF2023prl, GoLA2005prl}) or by chemostatting certain reactant species \cite{KiZw2021jotrsi, CoGA2022prl, Zwic2022cocis}. One could also render the interactions nonreciprocal by introducing driving force terms that cannot be derived from an energy functional \cite{TSJT2020pre, SaAG2020prx, FrWT2021pre, StJT2022sm, FrTh2023prl}.

To introduce the general gradient dynamics form we have first briefly reviewed how thin-film equations without chemical reactions can be written as gradient dynamics in the framework of linear nonequilibrium thermodynamics. Next, we have  reviewed how chemical reactions governed by detailed balance are brought in gradient dynamics form, however, this time employing nonlinear nonequilibrium thermodynamics. Then, we have combined the linear approach for transport processes  (and possibly simple transfer processes between phases) with the nonlinear approach for chemical reactions. The resulting rather general \reviewchange{model} may be employed to describe various phenomena that involve the physics of wetting for simple or complex liquids and chemical reactions that couple to interface dynamics.

In particular, we have illustrated the gradient dynamics form for chemical reactions first for a single reaction with mass action kinetics in a spatially homogeneous system. By re-expressing the reaction rates via the chemical potentials and introducing the principle of detailed balance from simple thermodynamic considerations, we have motivated a well-known form of mass action type kinetics. The approach corresponds to a gradient dynamics with nonlinear relations between the variations of the underlying energy functional and the thermodynamic fluxes. Active and sustained out-of-equilibrium dynamics can then be obtained by breaking the detailed balanced structure of the kinetics. This may, for instance, be achieved by chemostatting certain reactant species or by directly assuming different \reviewchange{rate functions} for the forward and backward reactions (e.g., for irreversible reactions). 

The gradient dynamics form retains its thermodynamic consistency even for nonideal systems.  This naturally leads to modifications of standard mass action reaction kinetics when nonideal contributions to the energy functional are present. In this way, a wide variety of cross-couplings between chemical reactions and fluid-mechanical phenomena may be captured. Here, we have focused on the combination of chemical reactions and wetting physics. We have demonstrated the usefulness of the approach at the example of two mesoscopic hydrodynamic models for shallow drops of partially wetting liquid on homogeneous rigid solid substrates -- describing reactive wetting and reactive surfactants.

First, we have considered active wetting: drops of solutions or suspensions where the suspended particles may adsorb from the drop onto the substrate or desorb from the substrate into the drop, thereby altering the substrate's wettability. In this case, one finds that the adsorption-desorption process is influenced by the proximity of the contact line due to interactions between the adsorbed particles and the substrate. At thermodynamic equilibrium, this leads to an additional interface between regions with different but respectively homogeneous adsorbate coverage that is located in the contact line region. When driving the system, e.g., by chemostatting the solute within the drop, the drop may perpetually move across the substrate while rendering the underlying \reviewchange{substrate} less wettable. Such drops driven by chemically sustained wettability gradients had already been experimentally and theoretically studied \cite{BaBM1994n,DoOn1995prl,BrGe1995crassi,Genn1998pa,LeLa2000jacs,LeKL2002pre,ThJB2004prl,SKYN2005pre,JoBT2005epje,SuMY2006ptps,Kita2006ptps,YBJZ2012sm,Arsc2016l}. Here, we recover these results with a model based on a gradient dynamics form where controlled chemostatting allows to quantify the `distance from a passive situation'.

Second, we have considered reactive surfactants: sessile drops of simple liquids are covered by two species of insoluble surfactant which engage in an autocatalytic conversion reaction. Bringing the surfactants on the free surface additionally in contact with a chemostat, i.e., a vast `surfactant bath',  a through-flow of energy and mass is introduced. Systems involving reactive surfactants had already been modeled in an \textit{ad hoc} manner, e.g., for liquid film flows in \cite{DaPi1984jcis,PTTK2007pf}. Here, employing the developed gradient dynamics description, we have found the emergence of a geometry-induced chemo-mechanical coupling resulting from the interdependence of the autocatalytic reaction and the free surface geometry that is known to result in chemo-mechanical instabilities on strongly curved surfaces \cite{MJK2019prl, WuGF2023pre}. Note, however, that this effect vanishes when employing a long-wave approximation as a final step of the derivation, as we have done here. However, the reactive surfactant dynamics still couples to the liquid bulk within the drop via solutal Marangoni forces. Focusing on the chemostatted out-of-equilibrium case, we have used numerical path continuation and time simulations to determine the rather complex bifurcation structure that shows how increasingly complex modes of motion emerge with increasing chemostatic driving strength. Described examples include steady localized Turing-like patterns of surfactants, self-excited breathing and swaying modes of standing wave-like drop oscillations as well as stationary and time-periodic self-propelled drop motion. The latter represents a modulated traveling wave-like behavior and the observed periodic emergence and disappearance of advancing protrusions might be seen as a primitive model of cell propulsion where time-periodic protrusion dynamics can also be observed \cite{AHSR2023bj}. Further, it is remarkable that the  bifurcation structure shows features that are quite similar to the snaking of localized states described for the Swift-Hohenberg equation \cite{BuKn2006pre} and its massconserving equivalent, the phase-field-crystal (PFC) model \cite{TARG2013pre}. However, in contrast to these cases, here, the localization is caused by the finite drop extension that limits the region where Turing patterns can develop. This should be further scrutinized in the future.

The presented general framework for the gradient dynamics-based description of reactive thin-film hydrodynamics can be easily adapted to many related situations and systems by incorporating other contributions to the energy such as interaction terms between surfactants (e.g., to account for surfactant phase transitions \cite{KGFT2012njp}) or by adapting the mobilities to account for other transport channels. In the case of our second example, one could introduce film height-dependent \reviewchange{rate functions} [mobilities] in the nonconserved [conserved] part of the dynamics (\ref{eq:GD_autocatalytic_surfacants_h})-(\ref{eq:GD_autocatalytic_surfacants_g2}) to suppress surfactant reactions [diffusion] in the ultrathin adsorption layer outside the drop. Finally, although we have restricted ourselves to mesoscopic hydrodynamic models for the description of sessile drops, we stress that the presented gradient dynamics approach may also be employed in many other systems where chemical reactions and fluid flows interact.

\section*{Statements and Declarations}
\subsection*{Acknowledgements}
FV and UT thank Yutaka Sumino for valuable discussions on reactive wetting, FV further thanks Jan Diekmann and Christopher Henkel for helpful advice regarding the numerics shown in Section~\ref{sec:numerical_results_running_drops}.
\subsection*{Author contributions}
Conceptualization: FV, UT; Methodology: FV, UT; Formal analysis: FV; Software: FV; Investigation: FV; Visualization: FV; Interpretation: FV, UT; Writing - original draft preparation: FV, UT; Writing - review and editing: FV, UT; Supervision: UT; Funding acquisition: UT; Project administration: UT; 
\subsection*{Competing interests}
The authors declare no competing interests.
\subsection*{Data Availability Statement}
The numerical data underlying this study and the python/gnuplot scripts used to present them are publicly available at Zenodo~\cite{VoTh2024Zenodo}.

\appendix
\renewcommand{\thesection}{\Alph{section}}
\renewcommand{\thesubsection}{\thesection.\arabic{subsection}}
\renewcommand{\thesubsubsection}{\thesubsection.\arabic{subsubsection}}

\section{\reviewchange{Supplemental calculations}}
\subsection{Gradient dynamics form of the reduced two-field model for $\mu_{a,0}=0$}\label{app:GD_proof_reduced_model_mu_a_0}
Here we show that the reduced model (\ref{eq:GD_mixture_reduced_h})-(\ref{eq:GD_mixture_reduced_b}) is a gradient dynamics for $\mu_{a,0}=0$ as stated in Section~\ref{sec:mixture_reduced_two_field_model}. First, we introduce the (reduced) free energy 
\begin{equation}
F = \int_\Omega\left[f(h, b)+\frac{\gamma}{2}\vert\nabla h\vert^2+g_b(b)+\frac{\sigma_b}{2}\vert\nabla b\vert^2\right]\text{d}^2 r \label{eq:F_reduced}
\end{equation}
and use it to obtain the liquid pressure and the chemical potential of $B$ as
\begin{align}
p = \frac{\delta F}{\delta h}&= \partial_h f - \gamma\Delta h,\\
\mu_b = \frac{\delta F}{\delta b} &= g_b'+\partial_b f -\sigma_b \Delta b.
\end{align}
We can then re-express (\ref{eq:GD_mixture_reduced_h})-(\ref{eq:GD_mixture_reduced_b}) as
\begin{align}
\partial_t h &= \nabla\cdot\left[Q_{hh}\nabla \frac{\delta F }{\delta h}\right],\label{eq:GD_mixture_reduced_h_with_Q}\\
\partial_t b &= \nabla\cdot\left[Q_{bb} \nabla \frac{\delta F }{\delta b}\right]-J_r,\label{eq:GD_mixture_reduced_b_with_Q}
\end{align}
where 
\begin{equation}
\tens{Q} = \left(\begin{array}{cc}
Q_{hh} & 0 \\ 
0 & Q_{bb}
\end{array}\right)=\left(\begin{array}{cc}
\frac{h^3}{3\eta} & 0 \\ 
0 & D_b b
\end{array}\right)
\end{equation}
is again a symmetric, positive (semi-)definite matrix. Therefore, the transport processes only lead to a reduction of the free energy. For the chemical reaction with $\mu_{a,0}=0$, we have the reactive flux
\begin{equation}
J_r = r(h) \left[\exp\left(\frac{\delta F /\delta b}{k_b T}\right)-1\right] = r(h)\left[X - 1\right],\label{eq:definition_J_appendix}
\end{equation}
where $X = \exp\left(\frac{\delta F /\delta b}{k_b T}\right)$. We now show that this flux is purely dissipative. To this end, we consider the change in $F$ due to reactive processes
\begin{align}
\left(\frac{\text{d}F }{\text{d}t}\right)_\text{react} &= \int_\Omega \frac{\delta F }{\delta b}\left(-J_r\right)\text{d}^2 r\\
&=-k_b T\int_\Omega r(h)(\ln X) \left(X-1\right)\text{d}^2 r\\
&\leq 0.
\end{align}
Here, we have only used the definitions of $J_r$ and $X$. The final inequality follows from the inequality $(\ln X)(X-1)\geq 0$. We note that this inequality does not hold for the more general expression $(\ln X)(X-X_0)$, i.e., for $\mu_{a,0}\neq 0$. Therefore, as thermodynamics suggests, Eqs.~(\ref{eq:GD_mixture_reduced_h})-(\ref{eq:GD_mixture_reduced_b}) only represent a gradient dynamics for $\mu_{a,0}=0$.
\subsection{Computation of energetic influx and dissipation for $\mu_{a,0}\neq 0$}\label{app:energetic_influx_dissipation_active}
In Section~\ref{sec:numerical_results_running_drops} we have discussed the rates of dissipation and of energetic influx from the chemostat for $\mu_{a,0}\neq 0$ in the reduced model (\ref{eq:GD_mixture_reduced_h})-(\ref{eq:GD_mixture_reduced_b}). Here, we briefly show how to compute these quantities by comparison with the passive case $\mu_{a,0}=0$. To this end, we compute the total change in free energy due to reactions for $\mu_{a,0}\neq 0$
\begin{flalign}
\left(\frac{\text{d}F}{\text{d}t}\right)_\text{react} &= \int_\Omega \frac{\delta F }{\delta b}\left(-J_r\right)\text{d}^2 r &&\\
&=-k_b T\int_\Omega r(h)\ln X  \left[X-\exp\left(\frac{\mu_{a,0}}{k_b T}\right)\right]\text{d}^2 r &&\\
&=\underbrace{-k_b T\int_\Omega r(h)\ln X  \left[X-1\right]\text{d}^2 r}_{\leq 0}+\underbrace{k_b T\int_\Omega r(h)\ln X \left[\exp\left(\frac{\mu_{a,0}}{k_b T}\right)-1\right]\text{d}^2 r}_{=R_\text{chem}}.&&\raisetag{2.2\baselineskip} \label{eq:split_dF_dt_passive_active}
\end{flalign}
Here, we have again used $\reviewchange{X=\exp\left(\frac{\delta F/\delta b}{k_b T}\right)}$ and the definition (\ref{eq:definition_J_appendix}) for $J_r$. In line (\ref{eq:split_dF_dt_passive_active}), we have separated the purely dissipative contributions \reviewchange{known} from the passive case in \reviewchange{Appendix~\ref{app:GD_proof_reduced_model_mu_a_0}}. We then \reviewchange{identify}
\begin{equation}
R_\text{chem} = k_b T\int_\Omega r(h)\ln X \left[\exp\left(\frac{\mu_{a,0}}{k_b T}-1\right)\right]\text{d}^2 r.
\end{equation}
Note that $R_\text{chem}$ may be positive or negative, i.e., the chemostat may be an energetic source or sink. If one chooses $\mu_{a,0}>0$, however, we usually find $R_\text{chem}>0$ after brief initial transients since then there is an energetic cost associated with replacing particles of species $a$ that adsorb onto the substrate. We now obtain the dissipation rate $R_\text{diss}$ from $\text{d}F/\text{d}t = -R_\text{diss}+R_\text{chem}$, where $\text{d}F/\text{d}t$ is the total derivative of $F$. Note that we have $R_\text{diss}\geq 0$.
\subsection{Details on nondimensionalization}
\subsubsection{Drops driven by chemically sustained wettability gradients
}\label{app:chemically_sustained_wettability_gradients_nondimensional}
We here briefly describe the derivation of the nondimensional version of Eqs.~(\ref{eq:GD_mixture_reduced_h})-(\ref{eq:GD_mixture_reduced_b}). For the nondimensionalization, we introduce the scales 
\begin{equation}
(x, y) = L (\tilde{x}, \tilde{y}) \hspace{0.5cm} h= l \tilde{h} \hspace{0.5cm} t = \tau\tilde{t} \hspace{0.5cm} b = b_0\tilde{b},
\end{equation}
with the scaling factors 
\begin{equation}
L = l\sqrt{\frac{\gamma}{k_b T b_0}} \hspace{0.5cm} l = h_a \hspace{0.5cm} \tau = \frac{L^2\eta}{h_a k_b T b_0}.
\end{equation}
Dimensionless quantities are denoted by tildes. The dimensionless time evolution equations then take the form
\begin{align}
\partial_{\tilde{t}} \tilde{h} &= \tilde{\nabla}\cdot\left[\frac{\tilde{h}^3}{3}\tilde{\nabla} \tilde{p}\right],\label{eq:GD_mixture_reduced_h_nondimensional}\\
\partial_{\tilde{t}} \tilde{b} &= \tilde{\nabla}\cdot\left[\tilde{D}_b \tilde{b} \tilde{\nabla}\tilde{\mu}_b\right]-\tilde{J}_r,\label{eq:GD_mixture_reduced_b_nondimensional}
\end{align}
with the nondimensional pressure and chemical potentials given as 
\begin{align}
\tilde{p} &= W\cdot\left(1+\lambda \tilde{b}\right)\cdot\left(\frac{1}{\tilde{h}^3}-\frac{1}{\tilde{h}^6}\right)-\tilde{\Delta}\tilde{h}, \\
\tilde{\mu}_b &= \ln\tilde{b}+\lambda W \left(-\frac{1}{2\tilde{h}^2}+\frac{1}{5\tilde{h}^5}\right)-\tilde{\sigma}_b\tilde{\Delta}\tilde{b}.
\end{align}
and with
\begin{equation}
\tilde{J}_r = \tilde{r}(\tilde{h})\left[\exp\left(\tilde{\mu}_b\right)-\exp\left(\tilde{\mu}_{a, 0}\right)\right],
\end{equation}
where the \reviewchange{rate function} is given by
\begin{equation}
\tilde{r}(\tilde{h}) = \tilde{r}_0 \xi(\tilde{h}).
\end{equation}
The remaining dimensionless quantities are 
\begin{align}
W &= \frac{A_H}{h_a^2 b_0 k_b T}, \\
\tilde{\sigma}_b &= \frac{b_0}{L^2 k_b T}\sigma_b,\\
\tilde{D}_b &= \frac{\tau k_b T}{L^2}D_b, \\
\tilde{r}_0 &= \frac{\tau}{b_0}r_0,\\
\tilde{\mu}_{a,0} &= \frac{1}{k_b T} \mu_{a,0},
\end{align}
where the wettability parameter $W$ describes the ratio of wetting energy to entropic substrate adsorbate contributions. In the main text, tildes denoting nondimensional quantities are omitted.
\subsubsection{Sessile drops covered by insoluble autocatalytic surfactants}\label{app:insoluble_autocatalytic_surfactants_nondimensional}
We here briefly summarize the nondimensionalization of Eqs.~(\ref{eq:h_hydrodynamic_autocatalysis})-(\ref{eq:g2_hydrodynamic_autocatalysis}). To this end, we introduce the scales
\begin{equation}
t = \tau \tilde{t}, \hspace{1cm} (x,y) = L (\tilde{x},\tilde{y}), \hspace{1cm} h = l\tilde{h}, \hspace{1cm} \Gamma_\alpha = a_1a_2\tilde{\Gamma}_\alpha\hspace{1cm} (f,g_s) = \kappa(\tilde{f},\tilde{g}_s),
\end{equation}
where dimensionless quantities are denoted by tildes. The respective scales are chosen as
\begin{equation}
\tau = \frac{L^2\eta}{\kappa l}, \hspace{1cm} L =\sqrt{\frac{\gamma^0}{\kappa}}l, \hspace{1cm} l=h_a, \hspace{1cm} \kappa = \frac{k_b T}{a_1 a_2}.
\end{equation}
Note that for this particular choice of scales the long-wave approximation $L \gg l$ simplifies to $\gamma^0\gg\tfrac{k_b T}{a_1 a_2}$. Therefore, neglecting surfactant contributions to the capillary pressure, as done in (\ref{eq:p_simplified_autocatalytic_surfactants}), is consistent with the present scaling. The nondimensional equations are now given by
\begin{align}
\partial_{\tilde{t}} \tilde{h} =&\: \tilde{\nabla}\cdot \left[\frac{\tilde{h}^3}{3}\tilde{\nabla}\left[W\left(\frac{1}{\tilde{h}^3}-\frac{1}{\tilde{h}^6}\right)-\tilde{\Delta} \tilde{h}\right]\right]+\tilde{\nabla}\cdot\left[\frac{\tilde{h}^2}{2 }\tilde{\nabla}\tilde{\Gamma}_1\right]+\tilde{\nabla}\cdot\left[\frac{\tilde{h}^2}{2}\tilde{\nabla}\tilde{\Gamma}_2\right],
\label{eq:h_hydro_autocatal_nondim}\\[2ex]
\begin{split}
\partial_{\tilde{t}} \tilde{\Gamma}_1 =&\: \tilde{\nabla}\cdot \left[\frac{\tilde{h}^2\tilde{\Gamma}_1}{2}\tilde{\nabla}\left[W\left(\frac{1}{\tilde{h}^3}-\frac{1}{\tilde{h}^6}\right)-\tilde{\Delta} \tilde{h}\right]\right]+\tilde{\nabla}\cdot\left[\left(\tilde{h}\tilde{\Gamma}_1+\tilde{D}_1\right)\tilde{\nabla}\tilde{\Gamma}_1\right]+\tilde{\nabla}\cdot\left[\tilde{h}\tilde{\Gamma}_1\tilde{\nabla}\tilde{\Gamma}_2\right]\\
&+\tilde{r}\left[\delta\tilde{\Gamma}_1^2\tilde{\Gamma}_2-\delta^3\tilde{\Gamma}_1^3\right]-\tilde{\beta}_1\ln(\tilde{\Gamma}_1 \delta)-\tilde{\beta\mu},
\end{split} \label{eq:g1_hydro_autocatal_nondim}
\\[2ex]
\begin{split}
\partial_{\tilde{t}} \tilde{\Gamma}_2 =&\: \tilde{\nabla}\cdot \left[\frac{\tilde{h}^2\tilde{\Gamma}_2}{2}\tilde{\nabla}\left[W\left(\frac{1}{\tilde{h}^3}-\frac{1}{\tilde{h}^6}\right)-\tilde{\Delta} \tilde{h}\right]\right]+\tilde{\nabla}\cdot\left[\tilde{h}\tilde{\Gamma}_2\tilde{\nabla}\tilde{\Gamma}_1\right]+\tilde{\nabla}\cdot\left[\left(\tilde{h}\tilde{\Gamma}_2+\tilde{D}_2\right)\tilde{\Gamma}_2\right]\\
&-\tilde{r}\left[\delta\tilde{\Gamma}_1^2\tilde{\Gamma}_2-\delta^3\tilde{\Gamma}_1^3\right]-\tilde{\beta}_2\ln(\tilde{\Gamma}_2 \delta^{-1})+\tilde{\beta\mu}.
\end{split}\label{eq:g2_hydro_autocatal_nondim}
\end{align}
Here, the dimensionless wettability parameter
\begin{equation}
W = \frac{A_H a_1a_2}{h_a^2 k_b T}
\end{equation}
expresses the ratio of wetting energy to surfactant energy. The remaining dimensionless quantities are
\begin{align}
\delta &= \frac{a_1}{a_2},\\
\tilde{D}_\alpha &= \frac{\eta a_1 a_2}{h_a} D_\alpha, \\
\tilde{r} &= \tau a_1 a_2 r =\frac{\gamma^0 h_a\eta (a_1 a_2)^3}{(k_b T)^2} r,\\
\tilde{\beta}_\alpha &= \tau a_1 a_2 k_b T\beta_\alpha = \frac{\gamma^0 h_a\eta (a_1 a_2)^3}{k_b T}\beta_\alpha,\\
\tilde{\beta\mu} &= \tau a_1 a_2 \beta\mu =\frac{\gamma^0 h_a\eta (a_1 a_2)^3}{(k_b T)^2}\beta\mu.
\end{align}
In the main text, we omit tildes denoting dimensionless quantities.
\section{Numerical methods}\label{app:numerical_details}
Here, we briefly describe the numerical methods \reviewchange{employed} to obtain the results presented in Sections~\ref{sec:numerical_results_running_drops} and \ref{sec:numerical_results_surfactants}.
\subsection{Time simulations}
For \reviewchange{all} time simulations, we use the finite-element library \textsc{oomph-lib} \cite{HeHa2006} \reviewchange{with linear Lagrange elements} \reviewchange{and adaptive mesh refinement based on a Zienkiewicz–Zhu error estimator \cite{ZiZh1992ijfnmie,ZiZh1992ijfnmie2} on a one-dimensional periodic domain.} Typical numbers of elements are $\reviewchange{n_{el}}=5000$ (Section~\ref{sec:numerical_results_running_drops}) and $\reviewchange{n_{el}}=1000$ (Section~\ref{sec:numerical_results_surfactants}). Time-stepping is employed using a second-order backward differentiation scheme (BDF[2]) and is usually adaptive. An exception \reviewchange{are} the data shown in Figs.~\ref{fig:run_driven_plots}a,b which \reviewchange{are} obtained with a constant timestep of $\text{d}t=5$ to produce equidistant spacing \reviewchange{of the panels}.
\subsection{Path continuation}\label{app:continuation}
Given a system of (differential) equations, path continuation methods \cite{KrauskopfOsingaGalan-Vioque2007,DTZF2017n,SaNe2016epjt,DWCD2014ccp,DoKK1991ijbc,Govaerts2000} may be employed to determine entire branches of states (steady or time-periodic) as a control parameter is continuously varied. Importantly, path continuation methods work regardless of the stability of the state of interest, in contrast to time simulations which typically only result in linearly stable states. Branches can then be arranged in bifurcation diagrams such as Fig.~\ref{fig:drop_autocatalysis_bifdiag_betamu} that reveal the relations and transitions between the various states of the system. A simple approach to the continuation of steady states consists in starting with a known steady state and then slightly changing the control parameter. Using the state at the previous parameter value as initial data for a Newton method, the corresponding state can be computed at a new parameter value. By consecutively altering the continuation parameter and computing the corresponding state directly from the known state at a previous parameter value, one may obtain the corresponding branch. However, this simple procedure fails at critical points such as saddle-node bifurcations and in practice more sophisticated tangent predictor-Newton corrector methods are typically used (see e.g. \cite{Govaerts2000}). The continuation of time-periodic states follows the same principles but is computationally more costly.

To obtain entire bifurcation diagrams, it is often practical to start with a simple analytically or numerically known state. We use the state at thermodynamic equilibrium,~i.e., the state at $\beta\mu=0$ on the black branch in Fig.~\ref{fig:drop_autocatalysis_bifdiag_betamu}. Then numerical continuation is used to compute the entire branch and to simultaneously detect bifurcation points. At each of the latter one may switch branches to then follow the bifurcating branches (again detecting bifurcation points). Repeating the procedure, the entire bifurcation diagram is obtained step-by-step. However, the strategy may not work when states exist that are nested too deeply within the bifurcation structure or when (the often tricky) branch switching fails. Then, it may be useful to initialize continuation for part of the branches starting with states directly obtained from time simulations. Here, we do this for all branches of time-periodic states as branch switching fails at many of the occurring Hopf bifurcations. We stress that all branches in Fig.~\ref{fig:drop_autocatalysis_bifdiag_betamu} are computed using path continuation regardless of the particular initialization strategy. Note that additionally all stable states have also been recovered in time simulations.

To perform continuation of states described by Eqs.~(\ref{eq:h_hydrodynamic_autocatalysis})-(\ref{eq:g2_hydrodynamic_autocatalysis}), we discretize the model using the finite-element method with linear Lagrange elements on a one-dimensional periodic domain. We employ the continuation package \textsc{pde2path} \cite{UeWR2014nmma} which uses pseudo-arclength continuation with a predictor-corrector method. In particular, \textsc{pde2path} provides extensive methods for the treatment of partial differential equations with continuous symmetries such as volume conservation or translational invariance, which are both present in Eqs.~(\ref{eq:h_hydrodynamic_autocatalysis})-(\ref{eq:g2_hydrodynamic_autocatalysis}) with periodic boundaries. We use static meshes with a typical number of nodes $n_x=300$ and $n_t=40$ in space and time, respectively.

\end{document}